\RequirePackage[british]{babel}
\pdfoutput=1
\documentclass[reqno,a4paper,12pt]{amsart}
\usepackage[a4paper,hmargin=2cm,tmargin=3cm,bmargin=3cm]{geometry}
\usepackage[utf8x]{inputenc}
\usepackage[T1]{fontenc}
\usepackage{fouriernc}
\usepackage{amsmath,amssymb,amstext,amsthm,amscd,mathrsfs,eucal,bm,slashed}
\usepackage{graphicx,color}
\usepackage{array}
\usepackage{cite}
\usepackage{hyperref}
\hypersetup{%
  pdftitle   = {Rigid supersymmetry, conformal coupling and twistor spinors},
  pdfkeywords = {rigid supersymmetry, conformal coupling, twistor spinors},
  pdfauthor  = {Paul de Medeiros},
  pdfcreator = {\LaTeX\ with package \flqq hyperref\frqq}
}
\PrerenderUnicode{éÉ}
\newcommand{\half}{\tfrac12}

\newcommand{\cL}{\mathcal{L}}

\newcommand{\fg}{\mathfrak{g}}

\newcommand{\fB}{\mathfrak{B}}

\newcommand{\fd}{\mathfrak{d}}

\newcommand{\fF}{\mathfrak{F}}
\newcommand{\fG}{\mathfrak{G}}

\newcommand{\fh}{\mathfrak{h}}

\newcommand{\fk}{\mathfrak{k}}

\newcommand{\fso}{\mathfrak{so}}

\newcommand{\fusp}{\mathfrak{usp}}

\newcommand{\fsp}{\mathfrak{sp}}
\newcommand{\fsu}{\mathfrak{su}}
\newcommand{\fu}{\mathfrak{u}}

\newcommand{\Cl}{\mathrm{C}\ell}

\newcommand{\GL}{\mathrm{GL}}

\newcommand{\RR}{\mathbb{R}}
\newcommand{\CC}{\mathbb{C}}
\newcommand{\HH}{\mathbb{H}}

\newcommand{\eD}{\mathscr{D}}
\newcommand{\eH}{\mathscr{H}}

\newcommand{\eL}{\mathscr{L}}
\newcommand{\eM}{\mathscr{M}}
\newcommand{\eN}{\mathscr{N}}

\newcommand{\AdS}{\mathrm{AdS}}

\DeclareMathOperator{\Mat}{Mat}

\theoremstyle{plain}

\theoremstyle{definition}

\newcommand{\MUNCH}[1]{\relax}

\allowdisplaybreaks
\begin{document}
\title[Rigid supersymmetry, conformal coupling and twistor spinors]{Rigid supersymmetry, conformal coupling and twistor spinors}
\author[de Medeiros]{Paul de Medeiros}
\address{School of Mathematics, Cardiff University, Senghennydd Road, Cardiff CF24 4AG, Wales, UK}
\email{demedeirospf@cardiff.ac.uk}
\date{\today}
\vspace*{-.5cm}
\begin{abstract}
We investigate the relationship between conformal and spin structure on lorentzian manifolds and see how their compatibility influences the formulation of rigid supersymmetric field theories. In dimensions three, four, six and ten, we show that if the Dirac current associated with a generic spinor defines a null conformal Killing vector then the spinor must obey a twistor equation with respect to a certain connection with torsion. Of the theories we consider, those with classical superconformal symmetry in Minkowski space can be reformulated as rigid supersymmetric theories on any lorentzian manifold admitting twistor spinors. In dimensions six and ten, we also describe rigid supersymmetric gauge theories on bosonic minimally supersymmetric supergravity backgrounds. 
\end{abstract}
\maketitle
\vspace*{-.5cm}
{\small{\tableofcontents}}

\clearpage

\section{Introduction}
\label{sec:introduction}

Supersymmetry continues to provide a powerful framework for the precise understanding of many important structures in quantum field theory. Given a field theory with rigid supersymmetry in flat space, it is often possible to couple it to supergravity in such a way that it retains local supersymmetry in curved space. It is well-known that supersymmetric couplings of this kind can be induced holographically in a superconformal field theory in Minkowski space that is dual to a string theory in an asymptotically anti-de Sitter background. A goal which has attracted an increasing amount of attention in the recent literature is the characterisation of {\emph{rigid}} supersymmetry for field theories in curved space \cite{Festuccia:2011ws,Jia:2011hw,Samtleben:2012gy,Klare:2012gn,Dumitrescu:2012he,Cassani:2012ri,Liu:2012bi}. This is a natural question which has been addressed in several isolated cases some time ago (e.g. see \cite{Shuster:1999zf,Blau:2000xg} and references therein). 

Recent interest in this topic stems primarily from the impressive results obtained by Pestun in \cite{Pestun:2007rz} for Wilson loops in maximally supersymmetric Yang-Mills theory on $S^4$. The calculation was accomplished by using a rigid supersymmetry of the theory on $S^4$ to localise the path integral and reduce it to an exactly solvable matrix model. A key ingredient is the contribution from non-minimal curvature couplings that are required to realise rigid supersymmetry on $S^4$. The scalar curvature plays the r\^{o}le of an infrared regulator in correlation functions. This localisation technique has since been replicated in several other contexts to yield many more important exact results. For example, in \cite{Kapustin:2009kz,Drukker:2010nc} it was applied to superconformal Chern-Simons theories on $S^3$, heralding many breakthroughs in our understanding of M2-branes. More recently, it has also been used in the context of rigid supersymmetric gauge theories on $S^5$ \cite{Kallen:2012cs,Hosomichi:2012he,Kallen:2012va,Kim:2012av} to confirm a number of expected properties for M5-branes \cite{Douglas:2010iu,Lambert:2010iw}. 

The broad scope for potential applications of this localisation technique in a wider class of rigid supersymmetric field theories in curved space has motivated a more systematic review of their construction. The strategy advocated by Festuccia and Seiberg in \cite{Festuccia:2011ws} has led the way. In four dimensions, they described how a large class of rigid supersymmetric non-linear $\sigma$-models in curved space can be obtained by taking a decoupling limit (in which the Planck mass goes to infinity) of the corresponding locally supersymmetric theory coupled to minimal supergravity. In this limit, the gravity supermultiplet is effectively frozen out, leaving only the fixed bosonic supergravity fields as data encoding the geometry of the supersymmetric curved background. Following this paradigm, several other works \cite{Jia:2011hw,Samtleben:2012gy,Klare:2012gn,Dumitrescu:2012he,Cassani:2012ri,Liu:2012bi} have explored the structure of rigid supersymmetry for field theories in three and four dimensions on curved manifolds in both euclidean and lorentzian signature. Since the localisation technique described above is most straightforwardly applied on compact manifolds, the natural focus of much of this work has been on the riemannian case though the lorentzian case, which is more suited to holographic applications, is also treated specifically in \cite{Festuccia:2011ws,Jia:2011hw,Cassani:2012ri,Liu:2012bi}. 

For a supersymmetric field theory in four-dimensional Minkowski space, there is a well-defined recipe \cite{vanNieuwenhuizen:1996tv} for analytically continuing to euclidean signature in the path integral. For supermultiplets involving Majorana fermions in lorentzian signature, at the classical level, the naive Wick rotation to euclidean signature is compatible only with a complexified form of the original supersymmetry algebra, effectively doubling the number of fermionic degrees of freedom. The balance can be restored classically by performing another Wick rotation with respect to the R-symmetry to recover a different real form of the original superalgebra in euclidean signature. Alternatively, as in \cite{vanNieuwenhuizen:1996tv}, one can restore the balance at the quantum level by summing over only a chiral projection of the complexified fermions in the path integral. For rigid supersymmetric field theories in curved space though, the relationship between lorentzian and euclidean signature is typically less transparent and some critical distinctions in four dimensions were clarified in \cite{Cassani:2012ri,Liu:2012bi}.

The aim of this paper is to investigate the relationship between conformal and spin structure on lorentzian manifolds, and how their compatibility influences the formulation of rigid supersymmetric field theories. To this end, we will focus on some well-known minimal supermultiplets in spacetime dimensions three, four, six and ten. On a curved lorentzian spin manifold, a field theory with a rigid symmetry generated by at least one spinor supercharge will be referred to as being {\emph{minimally supersymmetric}}. This should be contrasted with the situation in Minkowski space where minimal supersymmetry means that the number of linearly independent supercharges equals the dimension of the spinor representation to which the supercharges belong. In dimensions three and four, we shall recover no more than what is already known in the literature. At least by way of review, this should help contextualise some of the new results we obtain in dimensions six and ten within a more general framework. If a field theory on a lorentzian spin manifold $\eM$ has a rigid supersymmetry generated by a spinor $\epsilon$ and its associated Dirac current $\xi$ generates a null conformal isometry of $\eM$ that is compatible with the spin structure, we will see that $\epsilon$ must obey a twistor spinor equation with respect to a certain connection with torsion, whose form is dictated by the isotropy group of $\epsilon$. We will investigate the integrability conditions resulting from this twistor spinor equation and identify maximally supersymmetric solutions in dimensions three, four and six. For theories with classical superconformal symmetry in Minkowski space, we show how their conformal coupling on a lorentzian spin manifold gives a rigid minimally supersymmetric field theory provided the supersymmetry parameter is a twistor spinor. In dimensions six and ten, we also describe the formulation of rigid minimally supersymmetric Yang-Mills theory on bosonic supersymmetric backgrounds of the corresponding minimal Poincar\'{e} supergravity theories. 

For the most part, the rigid supersymmetry we describe in dimensions three, four, six and ten is very much predicated on the special properties of spinors in lorentzian signature. It is possible that some of these results could be transplanted into euclidean signature though we will not investigate that possibility here. In conformity with much of the existing literature, we shall ignore total derivative terms which arise from varying lagrangians. A systematic treatment of consistent supersymmetric boundary conditions for field theories on the class of lorentzian spin manifolds of interest is well beyond the scope of the present work. It is also perhaps worth emphasising that our analysis will be entirely classical. In dimensions greater than four, the supersymmetric field theories we consider are of course not expected to be perturbatively renormalisable and should only be thought of as the low-energy effective description in cases which admit a consistent ultraviolet completion.

The organisation of this paper is as follows. In section~\ref{sec:cliffordspin} we describe conventions and review some useful properties and identities for Clifford algebras and spinor representations in lorentzian signature. Section \ref{sec:generalisedtwistorspinors} contains a brief review of twistor spinors on lorentzian manifolds, focussing on pertinent classification results together with several special cases and generalisations that will be used in later sections. Section~\ref{sec:rigidsusyandconformalcoupling} describes our field theory conventions and illustrates the relationship between rigid supersymmetry and conformal coupling using the simple example of a free scalar supermultiplet. We then proceed to apply these ideas in the construction of rigid minimally supersymmetric field theories in dimensions three, four, six and ten. Section~\ref{sec:3dSUSY} describes the construction of off-shell gauge and matter supermultiplets on lorentzian three-manifolds admitting a twistor spinor (or a Killing spinor in the case of the supersymmetric Yang-Mills lagrangian). Section~\ref{sec:4dSUSY} does likewise for off-shell Yang-Mills and matter supermultiplets on lorentzian four-manifolds admitting a twistor spinor. In this case, if the Dirac current associated with the supersymmetry parameter generates a conformal isometry of the four-manifold then the supersymmetry parameter must obey a twistor spinor equation with respect to the same superconnection derived from minimal conformal supergravity in four dimensions (a result which was first obtained in \cite{Cassani:2012ri}). In section~\ref{sec:6dSUSY}, we construct off-shell Yang-Mills and on-shell matter and tensor supermultiplet couplings on lorentzian six-manifolds admitting a twistor spinor. We also describe how to accommodate a supersymmetric Yang-Mills lagrangian for a special type of algebraic twistor spinor in six dimensions. Furthermore, we construct non-minimal couplings which define the off-shell Yang-Mills supermultiplet on any bosonic supersymmetric background of minimal  Poincar\'{e} supergravity in six dimensions. In section~\ref{sec:10dSUSY}, we construct the on-shell Yang-Mills supermultiplet on any bosonic supersymmetric background of minimal  Poincar\'{e} supergravity in ten dimensions. We also comment on the obstructions to formulating Berkovits' (partially) off-shell version of this theory on a ten-dimensional lorentzian spin manifold when the rigid supersymmetry is generated by generic twistor spinors. Sections~\ref{sec:6dsugrabackgrounds} and \ref{sec:sugrabackgrounds} contain a brief synopsis of the classification of bosonic supersymmetric solutions of minimal Poincar\'{e} supergravity in dimensions six and ten. Finally, in section~\ref{sec:DecouplingCM}, we conclude with a brief summary of how the aforementioned rigid minimally supersymmetric Yang-Mills multiplet in ten dimensions follows from a particular decoupling limit of the locally supersymmetric Chapline-Manton theory.

\section{Clifford algebra and spinor conventions on lorentzian manifolds}
\label{sec:cliffordspin}

We take Minkowski space $\RR^{1,{\sf d}-1}$ to be equipped with mostly plus signature flat metric $\eta_{\mu\nu} = {\mbox{diag}} (-1,+1,...,+1)$ and orientation tensor $\varepsilon_{01...{\sf d}-1} = - \varepsilon^{01...{\sf d}-1} = \pm 1$. It is worth noting the identity $\varepsilon_{\mu_1 ... \mu_k \rho_{k+1} ... \rho_{\sf d}} \varepsilon^{\nu_1 ... \nu_k \rho_{k+1} ... \rho_{\sf d}} = - k! ({\sf d}-k)! \, \delta_{[ \mu_1}^{\nu_1} ... \delta_{\mu_k ]}^{\nu_k}$, where indices are raised using the inverse metric $\eta^{\mu\nu}$.

The Clifford algebra $\Cl(1,{\sf d}-1)$ contains elements $\Gamma_\mu$ that are subject to the defining relation $\Gamma_\mu \Gamma_\nu + \Gamma_\nu \Gamma_\mu = 2\, \eta_{\mu\nu} {\bf 1}$. A convenient basis for the Clifford algebra is given in terms of elements $\Gamma_{\mu_1 ... \mu_k} = \Gamma_{[ \mu_1} ... \Gamma_{\mu_k ]}$, where square brackets denote skewsymmetrisation with weight one. Let us christen $\fd = {\lfloor \tfrac{{\sf d}}{2} \rfloor}$. Henceforth, it will be convenient to work with the $2^\fd$-dimensional irreducible representation of $\Cl(1,{\sf d}-1)$, wherein $\Gamma_\mu$ correspond to gamma matrices.

Some useful identities are   
\begin{align}\label{eq:GammaProducts}
\Gamma_\mu \Gamma_{\nu_1 ... \nu_k} &= \Gamma_{\mu \nu_1 ... \nu_k} + k\, \eta_{\mu \nu_1} \Gamma_{\nu_2 ... \nu_k} \nonumber \\
(-1)^k \, \Gamma_{\nu_1 ... \nu_k} \Gamma_\mu &= \Gamma_{\mu \nu_1 ... \nu_k} - k\, \eta_{\mu \nu_1} \Gamma_{\nu_2 ... \nu_k} \nonumber \\
& \nonumber \\
\Gamma_{\mu\nu} \Gamma_{\rho_1 ... \rho_k} &= \Gamma_{\mu\nu \rho_1 ... \rho_k} - 2k\, \eta_{\mu \rho_1} \Gamma_{\nu \rho_2 ... \rho_k} - k(k-1)\, \eta_{\mu \rho_1} \eta_{\nu \rho_2} \Gamma_{\rho_3 ... \rho_k} \\
\Gamma_{\rho_1 ... \rho_k} \Gamma_{\mu\nu} &= \Gamma_{\mu\nu \rho_1 ... \rho_k} + 2k\, \eta_{\mu \rho_1} \Gamma_{\nu \rho_2 ... \rho_k} - k(k-1)\, \eta_{\mu \rho_1} \eta_{\nu \rho_2} \Gamma_{\rho_3 ... \rho_k} \nonumber \\
& \nonumber \\
\Gamma_{\mu_1 ... \mu_k} \Gamma_{\nu_1 ... \nu_l} &= \sum_{m=0}^{{\mathrm{inf}}(k,l)} (-1)^{km} \sigma_m \; m! {k\choose m} {l \choose m} \; \eta_{\mu_1 \nu_1} ... \eta_{\mu_m \nu_m} \Gamma_{\mu_{m+1} ... \mu_k \nu_{m+1} ... \nu_l} \nonumber ~.
\end{align}
Obvious skewsymmetrisation of indices on the right hand side in \eqref{eq:GammaProducts} has been omitted and it is the sign function 
\begin{equation}\label{eq:sign}
\sigma_m = (-1)^{m(m+1)/2}~,
\end{equation}
with $\sigma_1 = \sigma_2 = -1$, $\sigma_3 = \sigma_4 = +1$ and $\sigma_m = \sigma_{m+4}$ that appears on the right hand side of the last expression in \eqref{eq:GammaProducts}. The lorentz subalgebra $\fso(1,{\sf d}-1) < \Cl(1,{\sf d}-1)$ is spanned by $M_{\mu\nu} = \half \Gamma_{\mu\nu}$, obeying $[ M_{\mu\nu} , M_{\rho\sigma} ] = - \eta_{\mu\rho} M_{\nu\sigma} + \eta_{\nu\rho} M_{\mu\sigma} + \eta_{\mu\sigma} M_{\nu\rho} - \eta_{\nu\sigma} M_{\mu\rho}$. 

For any $\Theta \in \bigwedge^k \RR^{1,{\sf d}-1}$, let $\slashed \Theta = \tfrac{1}{k!} \Theta^{\mu_1 ... \mu_k} \Gamma_{\mu_1 ... \mu_k} \in \Cl(1,{\sf d}-1)$ and $\Vert \Theta \Vert^2 = \tfrac{1}{k!} \Theta^{\mu_1 ... \mu_k} \Theta_{\mu_1 ... \mu_k}$.  Components of the Hodge dual ${*\Theta} \in \bigwedge^{{\sf d}-k} \RR^{1,{\sf d}-1}$ are ${*\Theta}_{\mu_1 ... \mu_{{\sf d}-k}} = \tfrac{1}{k!} \, \varepsilon_{\mu_1 ... \mu_{{\sf d}-k} \nu_{{\sf d}-k+1} ... \nu_{{\sf d}}} \Theta^{\nu_{{\sf d}-k+1} ... \nu_{{\sf d}}}$.

Involutions of the Clifford algebra are generated by $\Gamma_\mu \mapsto \sigma_{\bf X} \, {\bf X} \, \Gamma_\mu \, {\bf X}^{-1}$, where $\sigma_{\bf X}$ is a sign and ${\bf X} \in \GL ( 2^\fd )$ in the gamma matrix representation. Let us fix a basis in this representation such that the gamma matrices are unitary, i.e. $\Gamma_\mu^\dagger  = \Gamma^\mu = - \Gamma_0 \Gamma_\mu \Gamma^0$. Under transposition, $\Gamma_\mu^t = \sigma_{\bf C} \,  {\bf C}  \Gamma_\mu {\bf C}^{-1}$, in terms of charge conjugation matrix ${\bf C}$ obeying ${\bf C}^t = \sigma_\fd \, {\bf C}$. For ${\sf d}$ odd, $\sigma_{\bf C} = (-1)^\fd$ while for ${\sf d}$ even we can take $\sigma_{\bf C} =-1$. This representation is preserved by conjugating $\Gamma_\mu \mapsto {\bf U} \Gamma_\mu {\bf U}^\dagger$, together with ${\bf C} \mapsto {\bf U}^t {\bf C} {\bf U}$, for any unitary matrix ${\bf U} \in {\mathrm{U}}( 2^\fd )$. This invariance can be used to fix ${\bf C}^\dagger {\bf C} = {\bf 1}$. Under complex conjugation, $\Gamma_\mu^* = \sigma_{\bf B} \, {\bf B}  \Gamma_\mu {\bf B}^{-1}$, where ${\bf B}$ is proportional to $\Gamma_0^* {\bf C}$ and $\sigma_{\bf B} = - \sigma_{\bf C}$. By fixing the coefficient of proportionality such that ${\bf B}$ is unitary, it follows that ${\bf B}^* {\bf B} = \sigma_{\bf C} \sigma_\fd \, {\bf 1}$ and ${\bf B}^\dagger {\bf C} = \sigma_{\bf C} \sigma_\fd \,  \Gamma^0$. 

The complexified action of $\fso(1,{\sf d}-1) < \Cl(1,{\sf d}-1)$ in the $2^\fd$-dimensional representation defines the Dirac spinor representation. The Dirac conjugate $\psi^\dagger \Gamma^0$ of a (fermionic) Dirac spinor $\psi$ defines a lorentz-invariant hermitian inner product on $\CC^{2^\fd}$. 

The reality  condition $\psi^* = {\bf B} \psi$ can be imposed on a non-zero Dirac spinor $\psi$ only if ${\bf B}^* {\bf B} = {\bf 1}$. It defines the Majorana spinor representation. In this case, ${\bf B}$ corresponds to a real structure on $\CC^{2^\fd}$ and requires $\sigma_{\bf C} \sigma_\fd =1$ which occurs only in ${\sf d} = 1,2,3,4 \; {\mathrm{mod}} \; 8$. Let ${\overline \psi} = \psi^t {\bf{C}}$ denote the Majorana conjugate of $\psi$. When $\psi^* = {\bf B} \psi$, the Dirac and Majorana conjugates of $\psi$ are identical and the associated hermitian inner product on $\CC^{2^\fd}$ is compatible with the real structure ${\bf B}$ only if $\sigma_{\fd} =-1$. Whence, Majorana spinors exist only in ${\sf d} = 2,3,4 \; {\mathrm{mod}} \; 8$. Indeed these are precisely the dimensions in which $\Cl(1,{\sf d}-1)$ is isomorphic to a matrix algebra over $\RR$. For example, $\Cl(1,1) \cong \Mat_2 ( \RR )$, $\Cl(1,2) \cong \Mat_2 ( \RR ) \oplus \Mat_2 ( \RR )$ and $\Cl(1,3) \cong \Mat_4 ( \RR )$.

Let $\{ {\bf e}_A \}$ denote a basis on $\CC^{2}$, equipped with the canonical action of $\fusp(2) = \fu(2) \cap \fsp(2,\CC)$ which preserves a complex symplectic form with components $\varepsilon_{AB}$ (normalised such that $\varepsilon_{12} = 1$). Given a non-zero element $\psi = \psi^A \, {\bf e}_A$, built from a pair of complex Dirac spinors $( \psi^1 , \psi^2 )$ which transform in the defining representation of $\fusp(2)$, one may impose an alternative reality condition $( \psi^A )^* = \varepsilon_{AB} \, {\bf B} \psi^B$ only if ${\bf B}^* {\bf B} = -{\bf 1}$. It defines the symplectic Majorana spinor representation. In this case, ${\bf B}$ corresponds to a quaternionic structure on $\CC^{2^\fd}$ and requires $\sigma_{\bf C} \sigma_\fd =-1$ which occurs only in ${\sf d} = 5,6,7,8 \; {\mathrm{mod}} \; 8$. When $( \psi^A )^* = \varepsilon_{AB} \, {\bf B} \psi^B$, the Dirac conjugate of $\psi^A$ equals $\varepsilon_{AB} \, {\overline \psi}^B$ and the associated hermitian inner product on $\CC^{2^\fd}$ is compatible with the quaternionic structure ${\bf B}$ only if $\sigma_{\fd} =1$. Whence, symplectic Majorana spinors exist only in ${\sf d} = 6,7,8 \; {\mathrm{mod}} \; 8$. Indeed these are precisely the dimensions in which $\Cl(1,{\sf d}-1)$ is isomorphic to a matrix algebra over $\HH$. For example, $\Cl(1,5) \cong \Mat_4 ( \HH )$ , $\Cl(1,6) \cong \Mat_4 ( \HH ) \oplus \Mat_4 ( \HH )$ and $\Cl(1,7) \cong \Mat_8 ( \HH )$.

For ${\sf d}$ even, the chirality matrix $\Gamma \in \Cl(1,d-1)$ is defined such that $\Gamma^2 = {\bf 1}$ and $\Gamma \, \Gamma_\mu = - \Gamma_\mu \Gamma$. The element $\Gamma_0 \Gamma_1 ... \Gamma_{{\sf d}-1}$ squares to $- \sigma_{{\sf d}-1} \, {\bf 1}$ and anticommutes with $\Gamma_\mu$. One can therefore take $\Gamma = \pm \Gamma_0 \Gamma_1 ... \Gamma_{{\sf d}-1}$ in ${\sf d}=2  \; {\mathrm{mod}} \; 4$ and $\Gamma = \pm i \, \Gamma_0 \Gamma_1 ... \Gamma_{{\sf d}-1}$ in ${\sf d}=4  \; {\mathrm{mod}} \; 4$. One can define projection operators ${\bf{P}}_\pm = \half \left( {\bf{1}} \pm \Gamma \right)$ and the chiral projections $\psi_\pm = {\bf{P}}_\pm \psi$ of a Dirac spinor $\psi$ are Weyl spinors with $\pm$ chirality (i.e. $\Gamma \psi_\pm = \pm \psi_\pm$). Since $\Gamma^* = {\bf B} \, \Gamma \, {\bf B}^{-1}$ in ${\sf d}=2  \; {\mathrm{mod}} \; 4$ and $\Gamma^* = - {\bf B} \, \Gamma \, {\bf B}^{-1}$ in ${\sf d}=4  \; {\mathrm{mod}} \; 4$, a Dirac spinor can be simultaneously Majorana and Weyl only in ${\sf d}=2  \; {\mathrm{mod}} \; 8$ or simultaneously symplectic Majorana and Weyl only in ${\sf d}=6  \; {\mathrm{mod}} \; 8$.

Given a non-zero bosonic Majorana spinor $\epsilon$, ${\bf C} \Gamma_\mu$ is symmetric so the one-form bilinear $\xi_\mu = {\overline \epsilon} \Gamma_\mu \epsilon$ is non-zero. Similarly, given a non-zero bosonic symplectic Majorana spinor $\epsilon$, ${\bf C} \Gamma_\mu$ is skewsymmetric so the one-form bilinear $\xi_\mu = \varepsilon_{AB} \, {\overline \epsilon}^A \Gamma_\mu \epsilon^B$ is non-zero. In either case, $\xi$ is real (via the identity ${\bf B}^t {\bf C}^* {\bf B} = {\bf C}$) and will be referred to as the {\emph{Dirac current}} of $\epsilon$. In Minkowski space, the vector defined by a Dirac current is either timelike or null. Moreover, $\xi$ is null only if ${\slashed \xi} \epsilon =0$ and ${\overline \epsilon} \epsilon =0$. 

This is so if $\epsilon$ is Majorana in ${\sf d}=3,4$, symplectic Majorana-Weyl in ${\sf d}=6$ or Majorana-Weyl in ${\sf d}=10$. In precisely these four cases, the map $\pi : \epsilon \mapsto \xi$ has an interesting structure. The set of spinors $\epsilon$ with unit norm describes a $(2{\sf d} -5)$-sphere. The image of this sphere under $\pi$ describes a \lq celestial' $S^{{\sf d} -2} \subset \RR^{1,{\sf d}-1}$, for which the timelike component of $\xi$ is fixed. Indeed, for ${\sf d}=3,4,6,10$, the map defines a bundle $\pi : S^{2{\sf d} -5} \longrightarrow S^{{\sf d} -2}$, whose typical fibre is isomorphic to $S^{{\sf d}-3}$. By Adams' theorem, these four bundles are therefore naturally identified with the Hopf fibrations associated with the four division algebras $\RR$, $\CC$, $\HH$ and $\mathbb{O}$ (although the non-associative octonionic structure is not compatible with the real structure of Majorana-Weyl spinors in ${\sf d}=10$).  

On a general spin manifold $\eM$ with lorentzian metric $g$, we define an orthonormal frame bundle in terms of vielbeins $e_\mu^\alpha$, with $g_{\mu\nu} = e_\mu^\alpha e_\nu^\beta \, \eta_{\alpha\beta}$. This frame bundle is chosen such that it preserves the spin structure on $\eM$, i.e. at each point in $\eM$, the matrix $( e_\mu^\alpha ) \in {\mbox{Spin}} (1,{\sf d}-1 )$. The expressions described above on $\RR^{1,{\sf d}-1}$ can be extended to $\eM$ in the obvious manner with respect to this frame, using vielbeins to transform between general coordinate indices $\mu , \nu ,...$ and local lorentz indices $\alpha , \beta , ...\,$.  Partial derivatives on $\RR^{1,{\sf d}-1}$ become covariant derivatives with respect to the Levi-Civita connection $\nabla$ on $\eM$. The action of $\nabla$ on a spinor $\epsilon$ is given by $\nabla_\mu \epsilon = \partial_\mu \epsilon + \tfrac{1}{4} \omega_\mu^{\alpha\beta} \, \Gamma_{\alpha\beta} \,\epsilon$, where the spin connection $\omega_\mu^{\alpha\beta}$ is defined by $d e^\alpha + \omega^\alpha_{\;\;\; \beta} \wedge e^\beta =0$. Given a field theory in Minkowski space, implementing the extension above gives a covariant field theory on $\eM$ that we will refer to as being {\emph{minimally coupled}}.

At a point in $\eM$, let $H_\epsilon$ denote the stabilising lie subgroup of ${\mbox{Spin}} (1,{\sf d}-1 )$ transformations which preserve a given non-vanishing spinor $\epsilon$. Let $\fh_\epsilon < \fso(1,{\sf d}-1)$ denote the corresponding isotropy lie algebra of $\epsilon$ at that point. Now assume that $\epsilon$ defines a nowhere-vanishing section of the spinor bundle on $\eM$, which has the same stabiliser $H_\epsilon$ at each point in $\eM$. This defines a so-called {\emph{$G$-structure}} on $\eM$, with $G = H_\epsilon$. It allows one to reduce the structure group of the frame bundle on $\eM$ from ${\mbox{Spin}} (1,{\sf d}-1 )$ to $H_\epsilon$. There is a homogenous space ${\mbox{Spin}} (1,{\sf d}-1 ) \, /\, H_\epsilon$ of possible reductions. For a given reduction, it can be shown that there must exist a connection $\nabla^\prime$ on the reduced spinor bundle with respect to which $\epsilon$ is parallel. Furthermore, the difference between the action of $\nabla^\prime$ and $\nabla$ on $\epsilon$ is given by an algebraic element,  the {\emph{intrinsic torsion}} $\tau \in {T^* \eM} \otimes \fk_\epsilon$, where $\fk_\epsilon \cong \fso(1,{\sf d}-1) \, / \, \fh_\epsilon$.  See \cite{SalHol} for a pedagogical introduction to this construction and related concepts. We shall make use of these results in our subsequent analysis. 

\section{Twistor spinors on lorentzian manifolds}
\label{sec:generalisedtwistorspinors}

A spinor $\epsilon$ on $\eM$ is called a {\emph{conformal Killing (or twistor) spinor}} if it obeys $\nabla_\mu \epsilon = \tfrac{1}{\sf d} \, \Gamma_\mu {\slashed \nabla} \epsilon$. The Dirac current $\xi$ of a twistor spinor $\epsilon$ defines a conformal Killing vector, obeying $\nabla_\mu \xi_\nu + \nabla_\nu \xi_\mu = \tfrac{2}{\sf d} \, g_{\mu\nu} \nabla_\rho \xi^\rho$. We defer to \cite{Baum:2002,BL:2003} and references therein for a comprehensive introduction to twistor spinors on lorentzian manifolds. In Minkowski space, a twistor spinor is necessarily of the form $\epsilon = \epsilon_0 + x^\mu \Gamma_\mu \epsilon_1$, in terms of a pair of constant spinors $\epsilon_0$ and $\epsilon_1$. For a superconformal field theory in Minkowski space, a supersymmetry transformation with twistor spinor parameter $\epsilon$ gives the action of the Poincar\'{e} and conformal supercharges with constant parameters $\epsilon_0$ and $\epsilon_1$ respectively.  

Let $R^\alpha_{\;\;\; \beta} = d \omega^\alpha_{\;\;\; \beta} + \omega^\alpha_{\;\;\; \gamma} \wedge \omega^\gamma_{\;\;\; \beta}$ denote the Riemann curvature of the Levi-Civita connection on $\eM$. In a coordinate basis, the Weyl tensor is $W_{\mu\nu\rho\sigma} = R_{\mu\nu\rho\sigma} + g_{\mu\rho} K_{\nu\sigma} - g_{\nu\rho} K_{\mu\sigma} - g_{\mu\sigma} K_{\nu\rho} + g_{\nu\sigma} K_{\mu\rho}$, written in terms of the Schouten tensor $K_{\mu\nu} = \tfrac{1}{{\sf d} -2} \left( - R_{\mu\nu} + \tfrac{1}{2({\sf d} -1)} \, g_{\mu\nu} \, R \right)$ with Ricci tensor $R_{\mu\nu} = R_{\mu\rho\nu\sigma} g^{\rho\sigma}$ and scalar curvature $R = R_{\mu\nu} g^{\mu\nu}$. Integrability of a twistor spinor $\epsilon$ on $\eM$ constrains the Weyl tensor such that
\begin{equation}\label{eq:twistorint}
W_{\mu\nu\rho\sigma} \Gamma^{\rho\sigma} \epsilon = 0 \; , \quad\quad \tfrac{1}{\sf d} \, W_{\mu\nu\rho\sigma} \Gamma^{\rho\sigma} {\slashed \nabla} \epsilon =  C_{\mu\nu\rho} \Gamma^\rho \epsilon~,
\end{equation}
where $C_{\mu\nu\rho} = \nabla_{\mu} K_{\nu\rho} - \nabla_{\nu} K_{\mu\rho}$ is the Cotton-York tensor. The maximum number of linearly independent twistor spinors on $\eM$ is $2^{\fd +1}$. Furthermore, from proposition 3.2 in \cite{Baum:2002}, it follows that this bound is saturated locally only if $\eM$ is conformally flat, with vanishing Cotton-York tensor for all ${\sf d} \geq 3$. However, if $\eM$ is not simply connected, there can exist global obstructions. For example, while ${\RR}^{1,{\sf d}-2} \times S^1$ is locally isometric to $\RR^{1,{\sf d}-1}$, a local twistor spinor $\epsilon = \epsilon_0 + x^\mu \Gamma_\mu \epsilon_1$ on $\RR^{1,{\sf d}-2} \times S^1$ is only single-valued when $\epsilon_1 =0$.  

A {\emph{Killing spinor}} is a special type of twistor spinor. If ${\sf d}$ is odd, a Killing spinor $\epsilon$ obeys ${\slashed \nabla} \epsilon = \alpha\, \epsilon$, for some constant $\alpha$. If ${\sf d}$ is even, the chiral projections of a Killing spinor $\epsilon$ obey ${\slashed \nabla} \epsilon_\pm = \alpha_\pm \, \epsilon_\mp$, for some constants $\alpha_\pm$ that are related to each other by complex conjugation if $\epsilon$ is Majorana. The Dirac current $\xi$ of a Killing spinor $\epsilon$ defines a Killing vector, obeying $\nabla_\mu \xi_\nu + \nabla_\nu \xi_\mu = 0$. Riemannian manifolds admitting Killing spinors were classified in \cite{Baum:1989,Bar:1993} and the cone construction of \cite{Bar:1993} has been utilised for lorentzian manifolds admitting Killing spinors in \cite{ACGL:2009,Matveev:2009}.

A spinor that is {\emph{parallel}} with respect to $\nabla$, i.e. $\nabla_\mu \epsilon =0$, is a special type of Killing spinor (with Killing constant equal to zero). By definition, a chiral Killing spinor is necessarily parallel. In Minkowski space, a Killing spinor is necessarily $\partial$-parallel, whence constant. Lorentzian manifolds admitting parallel spinors can be classified according to the holonomy of $\nabla$. This method was pioneered in \cite{Bryant:2000,FigueroaO'Farrill:1999tx} and a recent survey of the classification can be found in \cite{Baum:2012}.

The classification in ${\sf d} \geq 3$ of all local conformal equivalence classes of lorentzian spin manifolds admitting generic twistor spinors without zeros was established by Leitner in \cite{Leitner:2005} (see also \cite{Baum:2008} for a nice  review). There are five different classes, all of which admit global solutions, with $\eM$ being locally conformally equivalent to either
\\ [.2cm]
1) The product of $\RR^{1,1}$ with a riemannian manifold admitting parallel spinors. \\ [.1cm]
2) A lorentzian Einstein-Sasaki manifold. \\ [.1cm]
3) The product of a lorentzian Einstein-Sasaki manifold with a riemannian manifold admitting Killing spinors. \\ [.1cm]
4) A Fefferman space. \\ [.1cm]
5) A Brinkmann wave admitting parallel spinors. \\ [.2cm]
In cases (1) and (5), each twistor spinor is parallel and the associated Dirac current is respectively timelike and null. In case (2), each twistor spinor is the sum of two Killing spinors and the associated Dirac current is timelike. In case (4), the Fefferman space is even-dimensional and admits two linearly independent twistor spinors. The associated Dirac current is a regular null Killing vector on $\eM$. Conversely, from proposition 4.4 in \cite{BL:2003}, if $\eM$ is indecomposable and admits a twistor spinor $\epsilon$ whose Dirac current $\xi$ is a regular null Killing vector then $\xi^\mu \xi^\nu R_{\mu\nu}$ is constant and non-negative. Case (4) follows if $\xi^\mu \xi^\nu R_{\mu\nu} > 0$ while case (5) follows if $\xi^\mu \xi^\nu R_{\mu\nu} =0$. 

A more general first order differential equation for $\epsilon$ is of the form $\nabla_\mu \epsilon = {\sf M}_\mu \epsilon$, where ${\sf M}_\mu$ is an algebraic operator which can be thought of as a $\Cl(1,{\sf d}-1)$-valued one-form on $\eM$. Spinors of this type often arise when $\eM$ is a bosonic supersymmetric background of some Poincar\'{e} supergravity theory (i.e. the defining condition comes from setting to zero the Poincar\'{e} supersymmetry variation of the gravitino and ${\sf M}_\mu$ typically involves various background fluxes). There can of course be further algebraic constraints on the supersymmetry parameter from setting to zero the supersymmetry variations of any lower spin fermionic supergravity fields, e.g. the dilatino. As was pioneered in \cite{Festuccia:2011ws} for non-linear $\sigma$-models in ${\sf d}=4$, an effective strategy for finding rigid supersymmetric field theories on curved spaces is by taking a certain global limit (where the Planck mass goes to infinity) of a locally supersymmetric field theory coupled to supergravity. Indeed, one can follow a similar path for field theories coupled to conformal supergravity. In that case, setting to zero the combined Poincar\'{e} and conformal supersymmetry variation of the gravitino gives a twistor spinor equation $\eD_\mu \epsilon = \tfrac{1}{\sf d} \, \Gamma_\mu {\slashed \eD} \epsilon$, with respect to a certain {\emph{superconnection}} $\eD_\mu = \nabla_\mu + {\sf N}_\mu$ involving a $\Cl(1,{\sf d}-1)$-valued one-form ${\sf N}_\mu$ that is specified by the fields appearing in the particular conformal supergravity background in question. For example, in ${\sf d}=4$, the minimal off-shell conformal supergravity theory contains a real abelian one-form $a_\mu$ and ${\sf N}_\mu$ is proportional to $i a_\mu \Gamma$ in the associated conformal supergravity background.

Of course, for a given nowhere-vanishing spinor $\epsilon$ with global isotropy group $H_\epsilon$, it is always possible to use the $H_\epsilon$-structure to adapt a local basis such that $\nabla_\mu \epsilon = \tau_\mu \epsilon$, in terms of the intrinsic torsion $\tau \in {T^* \eM} \otimes \fk_\epsilon$. Indeed, it is precisely the local identification of the action of $\tau_\mu$ and ${\sf M}_\mu$ on $\epsilon$ for bosonic supersymmetric supergravity backgrounds which often allows them to be classified. In our forthcoming analysis, we will employ a similar strategy to derive a twistor spinor equation in ${\sf d} =3,4,6,10$ for a rigid supersymmetry parameter $\epsilon$ involving a certain connection with torsion on the spinor bundle, assuming only that its associated Dirac current $\xi$ defines a null conformal Killing vector. In each case, the isotropy algebra $\fh_\epsilon \cong \fg_\epsilon \ltimes \RR^{{\sf d}-2}$, where $\fg_\epsilon < \fso({\sf d}-2)$ is the isotropy algebra of a non-zero spinor in $\RR^{{\sf d}-2}$. In ${\sf d} =3,4$, $\fg_\epsilon$ is trivial. In ${\sf d} = 6$, $\fg_\epsilon$ is isomorphic to $\fsu(2) < \fso(4)$. In ${\sf d} = 10$, $\fg_\epsilon$ is isomorphic to $\fso(7) < \fso(8)$. We will see that the associated connection $\eD_\mu = \nabla_\mu + {\sf t}_\mu$ with respect to which $\epsilon$ obeys a twistor spinor equation must have a rather specific kind of torsion ${\sf t} \in ( {T^* \eM} \otimes \fso({\sf d}-2) \, / \, \fg_\epsilon ) \oplus \fg_\epsilon \subset {T^* \eM} \otimes \fk_\epsilon$. 

\section{Rigid supersymmetry and conformal coupling}
\label{sec:rigidsusyandconformalcoupling}

It will be convenient to the take the supersymmetry parameter $\epsilon$ to be a commuting spinor in our forthcoming analysis. A supersymmetry transformation $\delta_\epsilon$ will therefore act as an odd derivation on fields in the supermultiplet. In a rigid supersymmetric field theory on $\eM$, off-shell closure of the supersymmetry algebra means that $\delta_\epsilon^2$ must act as an even derivation on fields in the supermultiplet which generates a symmetry of the theory. (On-shell closure means that the previous statement holds only up to equations of motion.) 
\footnote{It is perhaps worth spelling out precisely how the conventional supersymmetry algebra is recovered from this approach. Take $\epsilon = \kappa {\hat \epsilon}$, with $\kappa$ some constant fermionic scalar parameter multiplying fermionic spinor ${\hat \epsilon}$. Given any pair of commuting spinor parameters $\epsilon_1$ and $\epsilon_2$ for which the supersymmetry algebra closes, it must also close for any linear combination of $\epsilon_1$ and $\epsilon_2$. Closure of the polarisation $\delta_{\epsilon_1 + \epsilon_2}^2 - \delta_{\epsilon_1}^2 - \delta_{\epsilon_2}^2 = \delta_{\epsilon_1} \delta_{\epsilon_2} + \delta_{\epsilon_2} \delta_{\epsilon_1} = \kappa_1 \kappa_2 \, [ \delta_{{\hat \epsilon}_1} , \delta_{{\hat \epsilon}_2} ]$ is therefore guaranteed for the commutator of the associated conventional supersymmetry transformations $\delta_{{\hat \epsilon}_1}$ and $\delta_{{\hat \epsilon}_2}$, which act as even derivations on fields in the supermultiplet.} 
We will see that this even symmetry generically involves several contributions including a diffeomorphism which preserves the spin structure, an R-symmetry and a gauge transformation on $\eM$. It may also include a Weyl transformation if the supersymmetry transformations for the theory are Weyl-covariant. The vector $\xi$ defined by the Dirac current of $\epsilon$ parameterises the infinitesimal diffeomorphism and Weyl contributions respectively in terms of the lie derivative $\cL_\xi$ and the Weyl variation $\delta_\sigma$, with parameter $\sigma = -\tfrac{1}{{\sf d}} \nabla_\mu \xi^\mu$. The contribution from gauge variation $\delta_\Lambda$ comes with parameter $\Lambda = - \xi^\mu A_\mu$, where $A_\mu$ is the gauge field. Typically $\xi$ will correspond to a (conformal) Killing vector on $\eM$, in which case $\cL_\xi \psi = \xi^\mu \nabla_\mu \psi + \tfrac{1}{4} ( \nabla_\mu \xi_\nu ) \Gamma^{\mu\nu} \psi$, acting on any spinor $\psi$. This corresponds to the {\emph{spinorial lie derivative}} \cite{Kosmann:1972,BG:1992,Habermann:1996}, an operation which provides the natural (conformally) isometric action that is compatible with the spin structure on $\eM$. 

A minimally supersymmetric Yang-Mills multiplet in Minkowski space $\RR^{1,{\sf d}-1}$ contains only a bosonic gauge field $A_\mu$ and a fermionic gaugino $\lambda$. Both fields are valued in a real lie algebra $\fg$ with lie bracket $[-,-]$. To define a lagrangian, one also requires a $\fg$-invariant inner product on $\fg$, that will be written $(-,-)$. The gauge field has ${\sf d} -2$ real on-shell degrees of freedom. The number of real on-shell degrees of freedom of the gaugino is $2^\fd$ if it is symplectic Majorana, $2^{\fd -1}$ if it is Majorana, Weyl or symplectic Majorana-Weyl, or $2^{\fd -2}$ if it is Majorana-Weyl. The number of bosonic and fermionic degrees of freedom therefore only match in dimensions ${\sf d}=3,4,6,10$, where $\lambda$ must be respectively Majorana, Majorana (or Weyl), symplectic Majorana-Weyl, Majorana-Weyl. 

Let $\phi$ be a matter field valued in some representation $V$ of $\fg$. In terms of the action $\cdot$ of $\fg$ on $V$, we define $D_\mu \phi = \partial_\mu \phi + A_\mu \cdot \phi$. This transforms covariantly as $\delta_\Lambda ( D_\mu \phi ) = - \Lambda \cdot D_\mu \phi$ under infinitesimal gauge transformations $\delta_\Lambda A_\mu = D_\mu \Lambda$ and $\delta_\Lambda \phi = - \Lambda \cdot \phi$, for any $\fg$-valued function $\Lambda$. The action of $\fg$ on itself is defined by the adjoint representation so that $A_\mu \cdot \Lambda = [ A_\mu , \Lambda ]$. The field strength $F_{\mu\nu} = [ D_\mu , D_\nu ] = \partial_\mu A_\nu - \partial_\nu A_\mu + [ A_\mu , A_\nu ]$ transforms covariantly as $\delta_\Lambda F_{\mu\nu} = [ F_{\mu\nu} , \Lambda ]$. To define a lagrangian for matter fields, one typically also requires a symmetric $\fg$-invariant inner product on $V$, that will be written $\langle -,- \rangle$.

A classical field theory on $\RR^{1,{\sf d}-1}$, whose fields $\Phi$ have dimensions $\Delta_\Phi$, is scale-invariant if it is preserved by transforming each $\Phi \mapsto {\mbox{e}}^{\Delta_\Phi \, \sigma} \Phi$, for any constant $\sigma$. On $\RR^{1,{\sf d}-1}$, $\partial_\mu$ has dimension $1$ while $\eta_{\mu\nu}$ and $\Gamma_\mu$ are dimensionless. In a minimally supersymmetric Yang-Mills multiplet on $\RR^{1,{\sf d}-1}$ (with $\fg$ non-abelian), the gauge field $A_\mu$ has dimension $1$, the gaugino $\lambda$ has dimension $\tfrac{3}{2}$ and the supersymmetry parameter $\epsilon$ has dimension $-\half$. This ensures the supersymmetry transformations scale covariantly, though the integral of the supersymmetric Yang-Mills lagrangian is only scale-invariant in ${\sf d} =4$. A field theory on $\eM$ is Weyl-invariant if it is preserved by transforming each field $\Phi \mapsto {\mbox{e}}^{w_\Phi \, \sigma} \Phi$, for any function $\sigma$ on $\eM$ and for some assignment of constant weights $w_\Phi$ to fields $\Phi$. The infinitesimal action of this Weyl transformation will be written $\delta_\sigma \Phi = w_\Phi \sigma \Phi$. On $\eM$, it is conventional to assign $e_\mu^\alpha$ weight $1$, whence $g_{\mu\nu}$ has weight $2$ and $\Gamma_\mu$ has weight $1$. It is useful to note that
\begin{align}\label{eq:WeylDer}
{\slashed \nabla} &\mapsto {\mbox{e}}^{-\tfrac{({\sf d} +1)\sigma}{2}} \, {\slashed \nabla} \, {\mbox{e}}^{\tfrac{({\sf d} -1)\sigma}{2}}  \nonumber \\
\nabla_\mu - \tfrac{1}{\sf d} \, \Gamma_\mu {\slashed \nabla}  &\mapsto {\mbox{e}}^{\tfrac{\sigma}{2}} \, \left( \nabla_\mu - \tfrac{1}{\sf d} \, \Gamma_\mu {\slashed \nabla}  \right) \, {\mbox{e}}^{-\tfrac{\sigma}{2}} ~,
\end{align}
under a Weyl transformation on $\eM$, when acting on spinors. An immediate corollary of the second transformation above is that the defining equation for a twistor spinor $\epsilon$ is Weyl-invariant provided $\epsilon$ is assigned weight $\half$. 

In a scale-invariant field theory on $\RR^{1,{\sf d}-1}$, let $r_\Phi$ denote the tensorial rank of each field $\Phi$. The minimal coupling of any such theory on $\eM$ is invariant under {\emph{global}} Weyl transformations by assigning weights $w_\Phi = r_\Phi - \Delta_\Phi$. For example, $w_\Phi = - \Delta_\Phi$ if $\Phi$ is a scalar or spinor field while $w_\Phi = 0$ for a non-abelian gauge field $\Phi = A_\mu$. Of course, the minimally coupled theory will typically not be invariant under local Weyl transformations on $\eM$. However, at least if the original theory on $\RR^{1,{\sf d}-1}$ is not just scale but conformally invariant, it is often possible to add certain improvement terms to the minimally coupled theory (which vanish identically in Minkowski space) such that Weyl-invariance is realised on $\eM$. In cases where this is possible, we will refer to the resulting Weyl-invariant theory on $\eM$ as being {\emph{conformally coupled}}.   

The simplest well-known example of a non-trivial conformal coupling is for a free bosonic scalar field $\Phi$ on $\RR^{1,{\sf d}-1}$, with lagrangian proportional to $\langle \partial_\mu \Phi , \partial^\mu \Phi \rangle$. In this case, $\Delta_\Phi = \tfrac{{\sf d}}{2} -1$ but the integral of the minimally coupled lagrangian on $\eM$ is not Weyl-invariant with weight $w_\Phi = 1- \tfrac{{\sf d}}{2}$. The conformally coupled lagrangian on $\eM$ is proportional to
\begin{equation}\label{eq:conformallycoupledscalar}
\langle \nabla_\mu \Phi , \nabla^\mu \Phi \rangle + \frac{{\sf d} -2}{4({\sf d} -1)} \, R \langle \Phi , \Phi \rangle ~,
\end{equation}  
where $R$ is the scalar curvature of $\eM$. 

Now consider the conformal coupling of an on-shell non-interacting minimally supersymmetric matter multiplet on $\RR^{1,{\sf d}-1}$ in which the bosonic scalar $\Phi$ is paired with a fermionic spinor $\Psi$. Their free equations of motion are $\square\, \Phi =0$ and ${\slashed \partial} \Psi =0$. Schematically, the on-shell supersymmetry transformations on $\RR^{1,{\sf d}-1}$ are of the form
\begin{equation}\label{eq:freemattersusy}
\delta_\epsilon \Phi = {\overline \epsilon}^\prime \Psi \; , \quad\quad \delta_\epsilon \Psi = \Gamma^\mu \epsilon \, \partial_\mu \Phi~,
\end{equation} 
where the bosonic supersymmetry parameter $\epsilon$ is of the same spinorial type as $\Psi$ and $\epsilon^\prime = ( {\bf B} \epsilon )^*$. These transformations close provided $\delta_\epsilon^2 = \xi^\mu \partial_\mu$ on-shell, with real translation parameter $\xi^\mu = {\overline \epsilon}^\prime \Gamma^\mu \epsilon$. This requires the identity $\epsilon {\overline \epsilon}^\prime = \half {\slashed \xi}$ acting on $\Psi$. The integral on $\RR^{1,{\sf d}-1}$ of a lagrangian of the form $\langle \partial_\mu \Phi , \partial^\mu \Phi \rangle + \langle {\overline \Psi} , {\slashed \partial} \Psi \rangle$ is also invariant under \eqref{eq:freemattersusy}. Any such theory is, of course, scale-invariant with $\Delta_\Psi = \Delta_\Phi + \half$ and $\Delta_\epsilon = -\half$.   

Supermultiplets of this type exist in ${\sf d}=3,4,6$, provided $\Phi$ is valued in associative division algebra $\mathbb{A} = \RR , \CC , \HH$ and $\Psi$ is based on a Majorana, (chiral projected) Majorana, symplectic Majorana-Weyl spinor representation. This ensures there are precisely ${\mbox{dim}}_\RR \, \mathbb{A} = {\sf d} -2$ matching on-shell bosonic and fermionic degrees of freedom and that the spinorial representation of $\Psi$ is compatible with $\mathbb{A}$. The inner product $\langle -,- \rangle$ used to define the lagrangian corresponds to the real part of the canonical euclidean inner product on $\mathbb{A}$. 

The associated Weyl weights on $\eM$ are $w_\Phi = 1- \tfrac{{\sf d}}{2}$, $w_\Psi = \tfrac{1-{\sf d}}{2}$ and $w_\epsilon = \half$. The minimally coupled kinetic term for $\Psi$ in the lagrangian on $\eM$ is proportional to $\langle {\overline \Psi} , {\slashed \nabla} \Psi \rangle$. From \eqref{eq:WeylDer}, it follows that this term is also conformally coupled. However, the minimally coupled supersymmetry transformations \eqref{eq:freemattersusy} are not Weyl-covariant. This can be remedied by improving the fermion variation such that 
\begin{equation}\label{eq:conformallycoupledsusy}
\delta_\epsilon \Psi =  \Gamma^\mu \epsilon \, \nabla_\mu \Phi + \left( 1 - \tfrac{2}{{\sf d}} \right)  {\slashed \nabla} \epsilon \, \Phi~.
\end{equation} 
Of course, adding this improvement term does not guarantee closure of the supersymmetry algebra on $\eM$ and the conformally coupled lagrangian need not be supersymmetric. Remarkably though, one finds that both properties hold on any $\eM$ provided $\epsilon$ is a twistor spinor. On-shell closure of the supersymmetry algebra here is such that $\delta_\epsilon^2 = \cL_\xi + \delta_\sigma + \delta_\rho$, using the fermionic equation of motion ${\slashed \nabla} \Psi =0$, where $\delta_\sigma$ is a Weyl variation with parameter $\sigma = -\tfrac{1}{\sf d}\, \nabla_\mu \xi^\mu$ and $\delta_\rho$ is an R-symmetry variation with parameter $\rho$ proportional to ${\overline \epsilon}^\prime {\slashed \nabla} \epsilon  - {\mathrm{Re}} \, ( {\overline \epsilon}^\prime {\slashed \nabla} \epsilon )$, valued in the imaginary part of $\mathbb{A}$. 

It is worth noting that the description above can be generalised such that $\epsilon$ is taken to be a section of the tensor product bundle of $\mathbb{A}$-valued spinors over $\eM$. One can define a non-trivial connection on this bundle of the form $\eD_\mu  = \nabla_\mu + {\sf A}_\mu$, where the connection one-form ${\sf A}_\mu$ is $\mathbb{A}$-valued and acts via left multiplication on sections. One must define this connection projectively in order for it to be compatible with Weyl symmetry. The corresponding projective bundle in ${\sf d}=3,4,6$ is then identified with a Hopf fibration associated with the respective division algebra $\RR , \CC , \HH$. This allows one to define the rigid supermultiplet on a wider class of backgrounds with supersymmetry parameter $\epsilon$ obeying the twistor spinor equation $\eD_\mu \epsilon = \tfrac{1}{\sf d} \Gamma_\mu {\slashed \eD} \epsilon$. 

Although this example may appear relatively trivial, we will see that it is a useful illustration of the general strategy for putting an interacting superconformal field theory in Minkowski space on a curved lorentzian manifold admitting twistor spinors, to define a rigid minimally supersymmetric theory via improvement terms of the form \eqref{eq:conformallycoupledscalar} and \eqref{eq:conformallycoupledsusy}.  

Without the inclusion of additional compensator fields, a supersymmetric field theory that is not conformally invariant in Minkowski space clearly cannot be put on a lorentzian spin manifold whilst preserving the rigid supersymmetry associated with a generic twistor spinor parameter. If this were so then, in Minkowski space, the generic twistor spinor parameter would generate a superconformal symmetry, contradicting the original assumption. Supersymmetric Yang-Mills theory in ${\sf d} \neq 4$ is an obvious example. However, in ${\sf d} =3,6$, although the minimally supersymmetric Yang-Mills lagrangian is not scale-invariant in Minkowski space, the off-shell supersymmetry transformations are. We will see how these off-shell supersymmetry transformations can be made Weyl-covariant on $\eM$, via the conformal coupling procedure described above, and that they close off-shell to describe a rigid supersymmetry algebra provided the parameter is a twistor spinor. Of course, as predicted by the argument above, these supersymmetry transformations do not preserve the minimally coupled Yang-Mills lagrangian for generic twistor spinor parameter. For certain classes of algebraic twistor spinors though, we will show that special improvement terms can be added to the Yang-Mills lagrangian to make it supersymmetric. We will see how a certain scalar component in the twistor spinor equation plays the r\^{o}le of a conformal compensator in the Yang-Mills lagrangian. 

\section{${\sf d}=3$}
\label{sec:3dSUSY}

The clifford algebra $\Cl(1,2) \cong {\mbox{Mat}}_{2} ( \RR ) \oplus {\mbox{Mat}}_{2} ( \RR )$ and its action on $\RR^2 \oplus \RR^2$ defines the Dirac spinor representation. A Majorana spinor representation is defined by selecting one of the two ${\mbox{Mat}}_{2} ( \RR )$ factors acting on $\RR^2$. For concreteness, we could take $\Gamma_\mu = \{ \left( \begin{smallmatrix} 0&1\\ -1&0 \end{smallmatrix} \right) , \left( \begin{smallmatrix} 0&1\\ 1&0 \end{smallmatrix} \right) , \left( \begin{smallmatrix} 1&0\\ 0&-1 \end{smallmatrix} \right) \}$ and ${\bf C} = \Gamma^0 = \left( \begin{smallmatrix} 0&-1\\ 1&0 \end{smallmatrix} \right)$ acting on Majorana spinors. 

The identities in \eqref{eq:GammaProducts} imply
\begin{equation}\label{eq:3dgammaidentities}
\Gamma^\rho \Gamma_\rho = 3\, {\bf{1}} \; , \quad\quad \Gamma^\nu \Gamma_{\mu} \Gamma_\nu = - \Gamma_{\mu} \; , \quad\quad \Gamma_{\mu\nu} = \varepsilon_{\mu\nu\rho} \Gamma^\rho \; , \quad\quad \Gamma_{\mu\nu\rho} = \varepsilon_{\mu\nu\rho} {\bf 1}~,
\end{equation} 
with $\varepsilon_{012} =1$.

Since ${\bf C}^t = -{\bf C}$ and $\Gamma_\mu^t = - {\bf C} \Gamma_\mu {\bf C}^{-1}$, any two fermionic Majorana spinors $\psi$ and $\chi$ obey
\begin{equation}\label{eq:3dbilinears}
{\overline \chi} \psi = {\overline \psi} \chi \; , \quad\quad {\overline \chi} \Gamma_\mu \psi = - {\overline \psi} \Gamma_\mu \chi ~.
\end{equation}
(For either $\psi$ or $\chi$ bosonic, one just puts an overall minus sign in \eqref{eq:3dbilinears}.)  

For $\psi$ and $\chi$ fermionic, a useful Fierz identity is 
\begin{equation}\label{eq:3dfierz}
\psi \, {\overline \chi} = - \tfrac{1}{2} \left( ( {\overline \chi} \psi ) {\bf{1}} + ( {\overline \chi} \Gamma^\mu \psi ) \Gamma_\mu \right)~.
\end{equation}
(For either $\psi$ or $\chi$ bosonic, \eqref{eq:3dfierz} just acquires an extra minus sign.) 

For any non-zero bosonic Majorana spinor $\epsilon$, \eqref{eq:3dbilinears} implies that only the Dirac current bilinear $\xi_\mu = {\overline \epsilon} \Gamma_\mu \epsilon$ does not vanish identically. The Fierz identity \eqref{eq:3dfierz} for $\epsilon$ reads
\begin{equation}\label{eq:3dfierzepsilon}
\epsilon {\overline \epsilon} = \half {\slashed \xi}~,
\end{equation}
and a useful subsidiary identity is
\begin{equation}\label{eq:3dfierzepsilon2}
-\Gamma^{\mu\nu} \epsilon {\overline \epsilon} \, \Gamma_\nu =  \xi^\mu {\bf 1}~.
\end{equation}

The isotropy algebra $\fh_\epsilon$ of $\epsilon$ is isomorphic to $\RR < \fso(1,2)$, thus $\fk_\epsilon \cong \RR^2$. Relative to a null orthonormal basis $( {\bf e}_+ , {\bf e}_- , {\bf e}_1 )$ on $\RR^{1,2}$, with $\eta_{+-} = 1 = \eta_{11}$, let us fix $\xi = {\bf e}_+$ so that $\Gamma_+ \epsilon =0$ and $\Gamma_1 \epsilon = \epsilon$. Adapting this basis to a local frame on $\eM$ allows us to express 
\begin{equation}\label{eq:3dtorsion}
\nabla_\mu \epsilon = a_\mu \epsilon + b_\mu \Gamma_- \epsilon~,
\end{equation}
in terms of a pair of real one-forms $a$ and $b$ which comprise the intrinsic torsion associated with the $\RR$-structure. Demanding that $\xi$ be a conformal Killing vector fixes $a_+ = -2 b_-$, $a_- = 0 = b_+$ and $a_1 = b_-$. Consequently, \eqref{eq:3dtorsion} becomes $\nabla_\mu \epsilon = \Gamma_\mu ( a_1 \epsilon + \half a_+ \Gamma_- \epsilon )$ implying $\epsilon$ must be a generic twistor spinor. Furthermore, $\xi$ is a Killing vector only if $\epsilon$ is a Killing spinor, with $a_+ =0$.

\subsection{Gauge supermultiplet}
\label{sec:3dSYMoffshell} 

The minimal gauge supermultiplet in ${\sf d}=3$ contains a bosonic gauge field $A_\mu$ and a fermionic Majorana spinor $\lambda$. 

The supersymmetry transformations on $\RR^{1,2}$ are     
\begin{align}\label{eq:3dsusy}
\delta_\epsilon A_\mu &= {\overline \epsilon} \Gamma_\mu \lambda \nonumber \\
\delta_\epsilon \lambda &= - \tfrac{1}{2} F^{\mu\nu} \Gamma_{\mu\nu} \epsilon~,
\end{align}
where $\epsilon$ is a constant bosonic Majorana spinor.  Squaring \eqref{eq:3dsusy} gives $\delta_\epsilon^2 = \xi^\mu \partial_\mu + \delta_\Lambda$ identically with gauge parameter $\Lambda = -\xi^\mu A_\mu$ and thus the supersymmetry algebra closes off-shell. 

Up to boundary terms on $\RR^{1,2}$, the Yang-Mills
\begin{equation}\label{eq:3dSYMlag}
\eL_{\mathrm{SYM}} = -\tfrac{1}{4} ( F_{\mu\nu} , F^{\mu\nu} ) -\half ( {\overline \lambda} , {\slashed D} \lambda )~,
\end{equation}
and Chern-Simons
\begin{equation}\label{eq:3dCSlag}
\eL_{\mathrm{CS}} = - \varepsilon^{\mu\nu\rho} ( A_\mu , \partial_\nu A_\rho + \tfrac{1}{3} [ A_\nu , A_\rho ]  ) + ( {\overline \lambda} , \lambda )~,
\end{equation}
lagrangians are both invariant under \eqref{eq:3dsusy}. Clearly the integral of $\eL_{\mathrm{CS}}$ is scale-invariant while the integral of $\eL_{\mathrm{SYM}}$ is not.
 
Let us now consider the minimally coupled version of \eqref{eq:3dsusy} on a three-dimensional lorentzian spin manifold $\eM$. Notice that they are automatically Weyl-covariant on $\eM$. Squaring them gives  
\begin{align}\label{eq:3dsusysquared}
\delta_\epsilon^2 A_\mu &= - \xi^\nu F_{\mu\nu} = \cL_\xi A_\mu + D_\mu \Lambda \nonumber \\
\delta_\epsilon^2 \lambda &= \xi^\mu \nabla_\mu \lambda + [ \lambda , \Lambda ] + \half ( \nabla_\mu \xi^\mu ) \lambda + ( {\overline \epsilon} \nabla_\mu \epsilon ) \, \Gamma^\mu \lambda~,
\end{align}
where $\cL_\xi A_\mu = \xi^\nu \partial_\nu A_\mu + ( \partial_\mu \xi^\nu ) A_\nu$ denotes the lie derivative along $\xi$. The third term on the right hand side of $\delta_\epsilon^2 \lambda$ corresponds to a Weyl variation $\delta_\sigma$ with parameter $\sigma = -\tfrac{1}{3} \nabla_\mu \xi^\mu$ and $w_\lambda = -\tfrac{3}{2}$. If $\xi$ is a conformal Killing vector, off-shell closure in \eqref{eq:3dsusysquared} requires $\delta_\epsilon^2 = \cL_\xi + \delta_\sigma +\delta_\Lambda$, which is tantamount to imposing ${\overline \epsilon} \nabla_\mu \epsilon = \tfrac{1}{4} \varepsilon_{\mu\nu\rho} \nabla^\nu \xi^\rho$. This is satisfied only if $\epsilon$ is a twistor spinor.

The integral of the minimally coupled Chern-Simons lagrangian \eqref{eq:3dCSlag} on $\eM$ is Weyl-invariant and supersymmetric, for any choice of $\epsilon$. 

The integral of the minimally coupled Yang-Mills lagrangian \eqref{eq:3dSYMlag} on $\eM$ is not Weyl-invariant and obviously cannot be conformally coupled without additional compensator fields since it was not scale-invariant in Minkowski space. Up to boundary terms, one finds $\delta_\epsilon \eL_{\mathrm{SYM}} = \tfrac{1}{6} \, ( F^{\mu\nu} , {\overline \lambda} \Gamma_{\mu\nu} {\slashed \nabla} \epsilon )$ so the minimally coupled lagrangian is generically supersymmetric only if $\nabla_\mu \epsilon =0$. 

However, now let $\nabla_\mu \epsilon = \Gamma_\mu ( \alpha \epsilon + \beta^\nu \Gamma_{\nu} \epsilon )$, in terms of a real function $\alpha$ and a real one-form $\beta$. Integrability of this twistor spinor equation implies $\alpha$ is constant and $\beta$ is closed. If $\beta = - d \varphi$, for some function $\varphi$, one can define a supersymmetric lagrangian 
\begin{equation}\label{eq:3dSYMlagcurved}
{\mbox{e}}^{-2\, \varphi} \left[ -\tfrac{1}{4} ( F_{\mu\nu} , F^{\mu\nu} ) -\half ( {\overline \lambda} , {\slashed D} \lambda ) -\tfrac{\alpha}{2}  ( {\overline \lambda} , \lambda ) \right]~.
\end{equation}
The prefactor ${\mbox{e}}^{-2\, \varphi}$ in \eqref{eq:3dSYMlagcurved} plays the r\^{o}le of a conformal compensator while $\alpha$ couples to a mass term for the gaugino. Note that one can use a Weyl transformation to fix $\varphi =0$. In this case $\epsilon$ is a Killing spinor obeying $\nabla_\mu \epsilon = \alpha \Gamma_\mu \epsilon$. 

\subsection{Matter supermultiplet}
\label{sec:3dmatteroffshell} 

The minimal on-shell matter supermultiplet on $\RR^{1,2}$ contains a real bosonic scalar $\phi$ and and a fermionic Majorana spinor $\psi$. To match the $2$ off-shell fermionic degrees of freedom of $\psi$, the $1$ off-shell bosonic degree of freedom of $\phi$ can be supplemented by a real bosonic auxiliary scalar ${\mathrm{C}}$. All fields are valued in a real orthogonal representation $V$ of $\fg$. The $\fg$-invariant inner product $\langle -,- \rangle$ is a real symmetric $\RR$-bilinear form on $V$.

The supersymmetry transformations for the matter supermultiplet coupled to the minimal gauge supermultiplet on $\RR^{1,2}$ are     
\begin{align}\label{eq:3dsusymatter}
\delta_\epsilon \phi &= {\overline \epsilon} \psi \nonumber \\
\delta_\epsilon \psi &= ( {\slashed D} \phi ) \epsilon + {\mathrm{C}}\, \epsilon \nonumber \\
\delta_\epsilon {\mathrm{C}} &= {\overline \epsilon} \left( {\slashed D} \psi + \lambda \cdot \phi \right)~,
\end{align}
where $\epsilon$ is a constant bosonic Majorana spinor.  Squaring \eqref{eq:3dsusymatter} gives $\delta_\epsilon^2 = \xi^\mu \partial_\mu + \delta_\Lambda$ off-shell.  

Up to boundary terms on $\RR^{1,2}$, the matter
\begin{equation}\label{eq:3dMlag}
\eL_{\mathrm{M}} = -\half \langle D_\mu \phi , D^\mu \phi \rangle -\half \langle {\overline \psi} , {\slashed D} \psi \rangle + \half \langle {\mathrm{C}} , {\mathrm{C}} \rangle + \langle \phi , {\overline \lambda} \cdot \psi \rangle~,
\end{equation}
and superpotential
\begin{equation}\label{eq:3dSuperPotentiallag}
\eL_{\mathrm{W}} = - \langle {\mathrm{C}} , d W ( \phi ) \rangle + \half \langle {\overline \psi} , d^2 W ( \phi ) \psi \rangle~,
\end{equation}
lagrangians are both invariant under \eqref{eq:3dsusymatter}, where the superpotential $W$ can be any real $\fg$-invariant function on $V$. The matter lagrangian $\eL_{\mathrm{M}}$ is scale-invariant with $( \phi , \psi , {\mathrm{C}} )$ having dimensions $(\half,1,\tfrac{3}{2})$. This is only true of $\eL_{\mathrm{W}}$ if $W$ has dimension $4$, i.e. if it is a quartic function of $\phi$.

The conformal coupling of \eqref{eq:3dsusymatter} and \eqref{eq:3dMlag} on $\eM$ follows via the improvement terms described in \eqref{eq:conformallycoupledsusy} and \eqref{eq:conformallycoupledscalar}. Relative to their minimal coupling on $\eM$, this involves adding a term $\tfrac{1}{3} \, \phi {\slashed \nabla} \epsilon$ to the right hand side of $\delta_\epsilon \psi$ in \eqref{eq:3dsusymatter} and a term $- \tfrac{1}{16} \, R \langle \phi , \phi \rangle$ to $\eL_{\mathrm{M}}$. If $\epsilon$ is a twistor spinor on $\eM$, the conformally coupled supersymmetry transformations close off-shell with $\delta_\epsilon^2 = \cL_\xi + \delta_\sigma +\delta_\Lambda$, where $( \phi , \psi , {\mathrm{C}} )$ are assigned their canonical weights $(-\half,-1,-\tfrac{3}{2})$. Moreover, the integral of the conformally coupled matter lagrangian on $\eM$ is invariant under these supersymmetry transformations.

The integral of the minimally coupled superpotential lagrangian $\eL_{\mathrm{W}}$ on $\eM$ is, of course, only Weyl-invariant when $W$ is a quartic function. Indeed, it is only then that it is supersymmetric.  

\section{${\sf d}=4$}
\label{sec:4dSUSY}

The clifford algebra $\Cl(1,3) \cong {\mbox{Mat}}_{4} ( \RR )$ and its action on $\RR^4$ defines the Majorana spinor representation. The action of the complexified clifford algebra $\Cl(1,3)_\CC$ on $\CC^4$ defines the Dirac spinor representation. The $\pm 1$ eigenspaces of $\Gamma$ under the action of $\Cl(3,1)_\CC$ are isomorphic to $\CC^2$ and define the $\pm$ chirality Weyl spinor representations. The action of a subalgebra ${\mbox{Mat}}_{2} ( \CC ) < {\mbox{Mat}}_{4} ( \RR )$ on $\CC^2$ which commutes with the complex structure $i\Gamma$ defines the representation associated with the chiral projection of a Majorana spinor. 

The identities in \eqref{eq:GammaProducts} imply
\begin{equation}\label{eq:4dgammaidentities}
\Gamma^\rho \Gamma_\rho = 4\, {\bf{1}} \; , \quad\quad \Gamma^\rho \Gamma_{\mu} \Gamma_\rho = -2\, \Gamma_{\mu}  \; , \quad\quad \Gamma^\rho \Gamma_{\mu\nu} \Gamma_\rho = 0  ~,
\end{equation} 
and 
\begin{equation}\label{eq:4dgammadual}
\Gamma_{\mu\nu} = -\tfrac{i}{2} \varepsilon_{\mu\nu\rho\sigma} \Gamma^{\rho\sigma} \Gamma \; , \quad\quad \Gamma_{\mu\nu\rho} = -i \varepsilon_{\mu\nu\rho\sigma} \Gamma^{\sigma} \Gamma \; , \quad\quad \Gamma_{\mu\nu\rho\sigma}  = i \varepsilon_{\mu\nu\rho\sigma} \Gamma~,
\end{equation}
where $\varepsilon_{0123} =-1$ and $\Gamma = i \Gamma_0 \Gamma_1 \Gamma_2 \Gamma_3$.

The charge conjugation matrix obeys $( {\bf C} \Gamma_{\mu_1 ... \mu_k} )^t = - \sigma_k \, {\bf C} \Gamma_{\mu_1 ... \mu_k}$ and so ${\overline{\Gamma_{\mu_1 ... \mu_k} \epsilon}} = \sigma_k \, {\overline{\epsilon}} \Gamma_{\mu_1 ... \mu_k}$ for any Majorana spinor $\epsilon$. Thus any two fermionic Majorana spinors $\psi$ and $\chi$ obey
\begin{equation}\label{eq:4dbilnears}
{\overline \chi} \Gamma_{\mu_1 ... \mu_k} \psi = \sigma_k \, {\overline \psi} \Gamma_{\mu_1 ... \mu_k} \chi~,
\end{equation}
and a useful Fierz identity is 
\begin{equation}\label{eq:4dfierz}
\psi \, {\overline \chi} = - \tfrac{1}{4} \left( ( {\overline \chi} \psi ) {\bf{1}} + ( {\overline \chi} \Gamma^\mu \psi ) \Gamma_\mu -\half  ( {\overline \chi} \Gamma^{\mu\nu} \psi ) \Gamma_{\mu\nu} -  ( {\overline \chi} \Gamma^\mu \Gamma \psi ) \Gamma_\mu \Gamma +  ( {\overline \chi} \Gamma \psi ) \Gamma \right)~.
\end{equation}
(For either $\psi$ or $\chi$ bosonic, \eqref{eq:4dbilnears} and \eqref{eq:4dfierz} just acquire an extra minus sign.) 

Any two chiral projections $\psi_\pm$ and $\chi_\pm$ (with the same $\pm$ chirality) obey
\begin{equation}\label{eq:4devenbilnears}
{\overline \chi_\pm} \Gamma_{\mu_1 ... \mu_k} \psi_\pm = 0~,
\end{equation} 
for any $k$ odd, since ${\overline{{\bf P}_\pm \chi}} = {\overline{\chi}} \, {\bf P}_\pm$ and ${\bf P}_\pm \Gamma_\mu = \Gamma_\mu {\bf P}_\mp$. Moreover, note that \eqref{eq:4dgammadual} implies the complex two-form ${\overline \chi_\pm} \Gamma_{\mu\nu} \psi_\pm = \mp \tfrac{i}{2} \varepsilon_{\mu\nu\rho\sigma} \, {\overline \chi_\pm} \Gamma^{\rho\sigma} \psi_\pm$. The chiral projections of \eqref{eq:4dfierz} give
\begin{align}\label{eq:4dfierz2}
\psi_\pm \, {\overline \chi_\pm} &= - \tfrac{1}{2} \left( ( {\overline \chi_\pm} \psi_\pm ) -\tfrac{1}{4}  ( {\overline \chi_\pm} \Gamma^{\mu\nu} \psi_\pm ) \Gamma_{\mu\nu} \right) {\bf{P}}_\pm \nonumber \\
\psi_\pm \, {\overline \chi_\mp} &= - \half ( {\overline \chi_\mp} \Gamma^\mu \psi_\pm ) \Gamma_\mu {\bf{P}}_\mp~.
\end{align} 
This implies $\psi_\pm \, {\overline \psi_\pm} = -\half ( {\overline \psi_\pm} \psi_\pm ) {\bf{P}}_\pm$ and $\psi_\pm \, {\overline \psi_\mp} =0$, using \eqref{eq:4dbilnears}. Whence $( {\overline \psi_\pm} \psi_\pm ) \, \psi_\pm =0$. Furthermore $( {\overline \psi_+} \Gamma_\mu \psi_- ) ( {\overline \psi_+} \Gamma_\nu \psi_- ) = \half \, \eta_{\mu\nu} \, ( {\overline \psi_+} \psi_+ ) ( {\overline \psi_-} \psi_- )$.
 
For any non-zero bosonic Majorana spinor $\epsilon$, \eqref{eq:4dbilnears} implies that only the bilinears $\xi_\mu = {\overline \epsilon} \Gamma_\mu \epsilon$ and $\zeta_{\mu\nu} = {\overline \epsilon} \Gamma_{\mu\nu} \epsilon$ built from $\epsilon$ do not vanish identically. The Fierz identity \eqref{eq:4dfierz} for $\epsilon$ reads
\begin{equation}\label{eq:4dfierzepsilon}
\epsilon {\overline \epsilon} = \tfrac{1}{4} ( {\slashed \xi} - {\slashed \zeta} )~,
\end{equation}
and a useful subsidiary identity is
\begin{equation}\label{eq:4dfierzepsilon2}
-\Gamma^{\mu\nu} \epsilon {\overline \epsilon} \, \Gamma_\nu = \left(  \epsilon {\overline \epsilon} - \half {\slashed\xi} \right) \Gamma^\mu + \xi^\mu {\bf 1}~.
\end{equation}

The isotropy algebra $\fh_\epsilon$ of $\epsilon$ is isomorphic to $\RR^2 < \fso(1,3)$, thus $\fk_\epsilon \cong \RR^4$. Relative to a null orthonormal basis $( {\bf e}_+ , {\bf e}_- , {\bf e}_1 , {\bf e}_2 )$ on $\RR^{1,3}$, with $\eta_{+-} = 1 = \eta_{11} = \eta_{22}$, let us fix $\xi = {\bf e}_+$ so that $\Gamma_+ \epsilon =0$. Adapting this basis to a local frame on $\eM$ allows us to express 
\begin{equation}\label{eq:4dtorsion}
\nabla_\mu \epsilon_+ = \alpha_\mu \epsilon_+ + \beta_\mu \Gamma_1 \Gamma_- \epsilon_+~,
\end{equation}
in terms of a pair of complex one-forms $\alpha$ and $\beta$ which comprise the intrinsic torsion associated with the $\RR^2$-structure. The expression for the negative chirality projection of $\epsilon$ is related to \eqref{eq:4dtorsion} by complex conjugation. Notice that $\nabla_\mu \xi_\nu$ does not involve the imaginary part of $\alpha$. Demanding that $\xi$ be a conformal Killing vector fixes ${\mathrm{Re}}\, \alpha_+ = 2 \, {\mathrm{Re}}\, \beta_1$, ${\mathrm{Re}}\, \alpha_- = 0$, ${\mathrm{Re}}\, \alpha_1 = - {\mathrm{Re}}\, \beta_-$, ${\mathrm{Re}}\, \alpha_2 = {\mathrm{Im}}\, \beta_-$, $\beta_1 = i\beta_2$ and $\beta_+ = 0$. Consequently, from \eqref{eq:4dtorsion}, one derives the twistor spinor equation $\eD_\mu \epsilon = \tfrac{1}{4} \Gamma_\mu {\slashed \eD} \epsilon$ with respect to the connection $\eD_\mu = \nabla_\mu + i a_\mu \Gamma$, where $a_+ = - {\mathrm{Im}} \, \alpha_+ + 2\, {\mathrm{Re}} \, \beta_2$, $a_- = - {\mathrm{Im}} \, \alpha_-$, $a_1 = - {\mathrm{Im}} \, \alpha_1 - 2\, {\mathrm{Re}} \, \alpha_2$ and  $a_2 = - {\mathrm{Im}} \, \alpha_2 + 2\, {\mathrm{Re}} \, \alpha_1$. Remarkably, it is precisely the same superconnection that appears in minimal conformal supergravity in four dimensions! This nice characterisation was first observed and explained in \cite{Cassani:2012ri}.    

Integrability of the twistor spinor equation $\eD_\mu \epsilon = \tfrac{1}{4} \Gamma_\mu {\slashed \eD} \epsilon$ implies
\begin{equation}\label{eq:4dtwistorint}
\tfrac{1}{4} W_{\mu\nu\rho\sigma} \Gamma^{\rho\sigma} \epsilon = - \tfrac{i}{3} f_{\rho [\mu} \Gamma_{\nu ]}{}^\rho \Gamma \epsilon - \tfrac{i}{3} f_{\mu\nu} \Gamma \epsilon - \tfrac{1}{6} \varepsilon_{\mu\nu\rho\sigma} f^{\rho\sigma} \epsilon~,
\end{equation}
where $f=da$. A background $\eM$ is locally maximally supersymmetric only if $f=0$ with vanishing Weyl tensor. If $\eM$ is indecomposable and simply connected, up to local conformal equivalence, any such solution must be flat with $a=0$. 

\subsection{Yang-Mills supermultiplet}
\label{sec:4dSYMoffshell} 

The minimal on-shell Yang-Mills supermultiplet in ${\sf d}=4$ contains a bosonic gauge field $A_\mu$ and a fermionic Majorana spinor $\lambda$. To match the $4$ off-shell fermionic degrees of freedom of $\lambda$, the $3$ off-shell bosonic degrees of freedom of the gauge field can be supplemented by a single real bosonic auxiliary scalar ${\mathrm{D}}$. All fields are $\fg$-valued. 

The supersymmetry transformations on $\RR^{1,3}$ are     
\begin{align}\label{eq:4dsusy}
\delta_\epsilon A_\mu &= {\overline \epsilon} \Gamma_\mu \lambda \nonumber \\
\delta_\epsilon \lambda &= -\tfrac{1}{2} F^{\mu\nu} \Gamma_{\mu\nu} \epsilon -i\, {\mathrm{D}} \, \Gamma \epsilon \nonumber \\
\delta_\epsilon {\mathrm{D}} &= -i\,  {\overline \epsilon} \Gamma {\slashed D} \lambda~,
\end{align}
where $\epsilon$ is a constant bosonic Majorana spinor.  Squaring \eqref{eq:4dsusy} gives $\delta_\epsilon^2 = \xi^\mu \partial_\mu + \delta_\Lambda$ identically with gauge parameter $\Lambda = -\xi^\mu A_\mu$ and therefore the supersymmetry algebra closes off-shell. 

Up to boundary terms, the Yang-Mills lagrangian
\begin{equation}\label{eq:4dSYMlag}
\eL_{\mathrm{SYM}} = -\tfrac{1}{4} ( F_{\mu\nu} , F^{\mu\nu} ) -\half ( {\overline \lambda} , {\slashed D} \lambda ) + \tfrac{1}{2} ( {\mathrm{D}} , {\mathrm{D}} )~,
\end{equation}
is invariant under \eqref{eq:4dsusy}. This lagrangian is also scale-invariant, with $( A_\mu , \lambda , {\mathrm{D}}  )$ having dimensions $(1,\tfrac{3}{2},2)$. There is also a $\fu(1)$ R-symmetry, under which $( A_\mu , \lambda_\pm , {\mathrm{D}}  )$ have charges $(0,\pm 1,0)$. 

We now consider the minimally coupled version of \eqref{eq:4dsusy} on a four-dimensional lorentzian spin manifold $\eM$. Notice that they are automatically Weyl-covariant on $\eM$. Squaring them gives  
\begin{align}\label{eq:4dsusysquared}
\delta_\epsilon^2 A_\mu &= \cL_\xi A_\mu + D_\mu \Lambda \nonumber \\
\delta_\epsilon^2 \lambda &= \xi^\mu \nabla_\mu \lambda + [ \lambda , \Lambda ] - \Gamma^{\mu\nu} \epsilon ( \nabla_\mu {\overline \epsilon} ) \Gamma_\nu \lambda \nonumber \\
\delta_\epsilon^2  {\mathrm{D}} &= \cL_\xi  {\mathrm{D}} +  [ {\mathrm{D}} , \Lambda ] + \half ( \nabla_\mu \xi^\mu ) \, {\mathrm{D}} + \tfrac{i}{2} F_{\mu\nu} \, {\overline \epsilon} \Gamma \Gamma^\rho \Gamma^{\mu\nu} \nabla_\rho \epsilon ~.
\end{align}
The third term on the right hand side of $\delta_\epsilon^2 {\mathrm{D}}$ corresponds to a Weyl variation $\delta_\sigma$ with parameter $\sigma = -\tfrac{1}{4} \nabla_\mu \xi^\mu$ and $w_{\mathrm{D}} = -2$. If $\xi$ is a conformal Killing vector, off-shell closure in \eqref{eq:4dsusysquared} requires $\delta_\epsilon^2 = \cL_\xi + \delta_\sigma + \delta_\Lambda + \delta_\rho$, where $\delta_\rho \Phi = r \rho \, \Phi$ on a field $\Phi$ with R-charge $r$, in terms of some function $\rho$ on $\eM$. From \eqref{eq:4dsusysquared}, one can read off $\rho = \tfrac{3}{4} \, {\overline \epsilon} \Gamma {\slashed \nabla} \epsilon$. Closure is then equivalent to the conditions 
\begin{equation}\label{eq:4dsusyCKV}
3\, {\overline \epsilon} \nabla_\mu \epsilon = \half \nabla^\nu \zeta_{\mu\nu} \; , \quad\quad 3\, {\overline \epsilon} \Gamma \nabla_\mu \epsilon = -\tfrac{i}{4} \varepsilon_{\mu\nu\rho\sigma}  \nabla^\nu \zeta^{\rho\sigma} \; , \quad\quad \nabla_{[\mu} \xi_{\nu ]} = i \varepsilon_{\mu\nu\rho\sigma}  {\overline \epsilon} \Gamma \Gamma^\rho \nabla^\sigma \epsilon~.
\end{equation}
They are solved if $\epsilon$ is a twistor spinor. In that case, the integral of the minimal coupling of \eqref{eq:4dSYMlag} on $\eM$ is Weyl-invariant and supersymmetric. 

If $\epsilon$ is a twistor spinor with respect to the connection $\eD_\mu = \nabla_\mu + i a_\mu \Gamma$, the same picture emerges after gauging the $\fu(1)$ R-symmetry. This means that a local R-symmetry variation $\delta_\rho \Phi = r \rho \, \Phi$ is accompanied by the variation $\delta_\rho a_\mu = i \partial_\mu \rho$. Therefore $\eD_\mu \Phi = \nabla_\mu \Phi + i r a_\mu \Phi$ transforms covariantly under $\delta_\rho$. The R-symmetry gauging is tantamount to replacing $\nabla_\mu$ with $\eD_\mu$ in all the expressions above. One then takes the abelian gauge field $a_\mu$ to be a fixed background one-form with Weyl weight zero.

\subsection{Matter supermultiplet}
\label{sec:4dmatteroffshell} 

The minimal on-shell matter supermultiplet in ${\sf d}=4$ contains a complex bosonic scalar $\phi$ and a fermionic Majorana spinor $\psi$. To match the $4$ off-shell fermionic degrees of freedom of $\psi$, the $2$ off-shell bosonic degrees of freedom of $\phi$ can be supplemented by a complex bosonic auxiliary scalar ${\mathrm{F}}$. All fields are valued in a complex unitary representation $V$ of $\fg$. The $\fg$-invariant symmetric inner product $\langle -,- \rangle$ is the real part of a $\CC$-sesquilinear hermitian form on $V$.

The supersymmetry transformations for the matter supermultiplet coupled to the minimal gauge supermultiplet on $\RR^{1,3}$ are     
\begin{align}\label{eq:4dsusymatter}
\delta_\epsilon \phi &= {\overline \epsilon} \psi_+ \nonumber \\
\delta_\epsilon \psi_+ &= 2 \, ( {\slashed D} \phi ) \epsilon_- + 2\, {\mathrm{F}} \epsilon_+ \nonumber \\
\delta_\epsilon {\mathrm{F}} &= {\overline \epsilon} \left( {\slashed D} \psi_+ + 2\, \lambda_- \cdot \phi \right)~,
\end{align}
where $\epsilon$ is a constant bosonic Majorana spinor.  Squaring \eqref{eq:4dsusymatter} gives $\delta_\epsilon^2 = \xi^\mu \partial_\mu + \delta_\Lambda$ off-shell.

Up to boundary terms on $\RR^{1,3}$, the matter
\begin{equation}\label{eq:4dMlag}
\eL_{\mathrm{M}} = -\langle D_\mu \phi , D^\mu \phi \rangle -\half \langle {\overline \psi}_+ , {\slashed D} \psi_+ \rangle + \langle {\mathrm{F}} , {\mathrm{F}} \rangle + \langle \phi , i{\mathrm{D}} \cdot \phi \rangle - 2\, \langle {\overline \psi}_+ , \lambda \cdot \phi \rangle~,
\end{equation}
and superpotential
\begin{equation}\label{eq:4dSuperPotentiallag}
\eL_{\mathrm{W}} = 2\, \langle {\mathrm{F}} , \partial W ( \phi ) \rangle - \half \langle {\overline \psi}_+ , \partial^2 W ( \phi ) \psi_+ \rangle~,
\end{equation}
lagrangians are both invariant under \eqref{eq:4dsusymatter}, where the superpotential $W$ can be any $\fg$-invariant holomorphic function $W$ on $V$. The matter lagrangian $\eL_{\mathrm{M}}$ is classically scale-invariant with $( \phi , \psi_+ , {\mathrm{F}} )$ having dimensions $(1,\tfrac{3}{2},2)$. This is only true of $\eL_{\mathrm{W}}$ if $W$ has dimension $3$, i.e. it must be a cubic function of $\phi$.

The conformal coupling of \eqref{eq:4dsusymatter} and \eqref{eq:4dMlag} on $\eM$ follows via the improvement terms described in \eqref{eq:conformallycoupledsusy} and \eqref{eq:conformallycoupledscalar}. Relative to their minimal coupling on $\eM$, this involves adding a term $\phi {\slashed \nabla} \epsilon_-$ to the right hand side of $\delta_\epsilon \psi_+$ in \eqref{eq:4dsusymatter} and a term $- \tfrac{1}{6} \, R \langle \phi , \phi \rangle$ to $\eL_{\mathrm{M}}$. If $\epsilon$ is a twistor spinor on $\eM$, the conformally coupled supersymmetry transformations close off-shell with $\delta_\epsilon^2 = \cL_\xi + \delta_\sigma +\delta_\Lambda + \delta_\rho$, where $( \phi , \psi_+ , {\mathrm{C}} )$ are assigned their canonical weights $(-1,-\tfrac{3}{2},-2)$ with R-charges $( \tfrac{2}{3} , -\tfrac{1}{3} , -\tfrac{4}{3} )$. Furthermore, the integral of the conformally coupled matter lagrangian on $\eM$ is invariant under these supersymmetry transformations.

The integral of the minimally coupled superpotential lagrangian $\eL_{\mathrm{W}}$ on $\eM$ is, of course, only Weyl-invariant when $W$ is a cubic function. Indeed, it is only then that it is supersymmetric.  

If $\epsilon$ is a twistor spinor with respect to $\eD_\mu = \nabla_\mu + i a_\mu \Gamma$, the same story follows after gauging the R-symmetry and replacing $\nabla_\mu$ with $\eD_\mu = \nabla_\mu + i r a_\mu$ in all expressions. 

\section{${\sf d}=6$}
\label{sec:6dSUSY}

The clifford algebra $\Cl(1,5) \cong {\mbox{Mat}}_{4} ( \HH )$ and its action on $\HH^4$ defines the Dirac spinor representation. The $\pm 1$ eigenspaces of $\Gamma$ under the action of ${\mbox{Mat}}_{4} ( \HH )$ are isomorphic to $\HH^2$ and define the $\pm$ chirality Weyl spinor representations. It is convenient to think of $\HH \cong \CC^2$ here and represent a quaternionic spinor in terms of complex doublets, which transform under the auxiliary $\fusp(2)$ action that was described in section~\ref{sec:cliffordspin}, relative to a basis $\{ {\bf{e}}_A \}$ on $\CC^2$. Auxiliary indices will be raised and lowered using the $\fusp(2)$-invariant symplectic form, such that $u_A  = \varepsilon_{AB} \, u^B$ (and $u^A = u_B \, \varepsilon^{BA}$ via the identity $\varepsilon_{AC} \varepsilon^{BC} = \delta_A^B$) for any $u \in \CC^2$. A second rank symmetric tensor $w$ on $\CC^2$ obeying the reality condition $( w^{AB} )^* = \varepsilon_{AC} \varepsilon_{BD} w^{CD}$ corresponds to the adjoint representation of $\fusp(2)$.  

The identities in \eqref{eq:GammaProducts} imply
\begin{equation}\label{eq:6dgammaidentities}
\Gamma^\sigma \Gamma_\mu \Gamma_\sigma = -4\, \Gamma_\mu \; , \quad\quad \Gamma^\sigma \Gamma_{\mu\nu} \Gamma_\sigma = 2\, \Gamma_{\mu\nu}  \; , \quad\quad \Gamma^\sigma \Gamma_{\mu\nu\rho} \Gamma_\sigma = 0  ~,
\end{equation} 
and 
\begin{equation}\label{eq:6dgammadual}
\Gamma_{\mu_1 ... \mu_k} \Gamma = - \sigma_{k-1} \, \tfrac{1}{(6-k)!} \, \varepsilon_{\mu_1 ... \mu_k \nu_{k+1} ... \nu_{6}} \Gamma^{\nu_{k+1} ... \nu_{6}}~,
\end{equation}
where $\varepsilon_{01...5} =1$ and $\Gamma = \Gamma_0 \Gamma_1 ... \Gamma_5$.

The charge conjugation matrix obeys $( {\bf C} \Gamma_{\mu_1 ... \mu_k} )^t = \sigma_k \, {\bf C} \Gamma_{\mu_1 ... \mu_k}$ and so ${\overline{\Gamma_{\mu_1 ... \mu_k} \epsilon^A}} = \sigma_k \, {\overline \epsilon}^A \Gamma_{\mu_1 ... \mu_k}$ for any symplectic Majorana spinor $\epsilon$. Thus any two fermionic symplectic Majorana spinors $\psi$ and $\chi$ obey
\begin{equation}\label{eq:6dbilnears}
{\overline \chi}^A \Gamma_{\mu_1 ... \mu_k} \psi^B = - \sigma_k \, {\overline \psi}^B \Gamma_{\mu_1 ... \mu_k} \chi^A~.
\end{equation}
(For either $\psi$ or $\chi$ bosonic, one just adds an overall minus sign in \eqref{eq:6dbilnears}.) We define the contracted bilinear ${\overline \chi} \Gamma_{\mu_1 ... \mu_k} \psi = \varepsilon_{AB} \, {\overline \chi}^A \Gamma_{\mu_1 ... \mu_k} \psi^B$. 

Any two symplectic Majorana-Weyl spinors $\psi_\pm$ and $\chi_\pm$ (with the same $\pm$ chirality) obey the identity ${\overline \chi_\pm^A} \Gamma_{\mu_1 ... \mu_k} \psi_\pm^B =0$ for any even $k$. For $\psi_\pm$ and $\chi_\pm$ fermionic, two useful Fierz identities are 
\begin{align}\label{eq:6dfierz}
\psi_\pm^A \, {\overline \chi_\pm^B} &= - \tfrac{1}{4} \left( ( {\overline \chi_\pm^B} \Gamma^\mu \psi_\pm^A ) \Gamma_\mu - \tfrac{1}{12} ( {\overline \chi_\pm^B}\Gamma^{\mu\nu\rho} \psi_\pm^A ) \Gamma_{\mu\nu\rho} \right) {\bf P}_\mp \nonumber \\
\psi_\pm^A \, {\overline \chi_\mp^B} &= - \tfrac{1}{4} \left( ( {\overline \chi_\mp^B} \psi_\pm^A ) {\bf 1} - \tfrac{1}{2} ( {\overline \chi_\mp^B}\Gamma^{\mu\nu} \psi_\pm^A ) \Gamma_{\mu\nu} \right) {\bf P}_\pm~.
\end{align}
(For either $\psi_\pm$ or $\chi_\pm$ bosonic, \eqref{eq:6dfierz} just acquires an extra minus sign.) 

From \eqref{eq:6dgammadual}, $\tfrac{1}{6} \varepsilon_{\mu\nu\rho\alpha\beta\gamma} \Gamma^{\alpha\beta\gamma} \psi_\pm^A = \pm \Gamma_{\mu\nu\rho} \psi_\pm^A$ for any symplectic Majorana-Weyl spinor $\psi_\pm$. Whence, for any symplectic Majorana spinor $\chi$, the three-form defined by each bilinear $ {\overline \chi}^A \Gamma_{\mu\nu\rho} \psi_\pm^B$ is self-dual for $\psi_+$ with positive chirality and anti-self-dual for $\psi_-$ with negative chirality.

Any pair of (anti-)self-dual three-forms $X, Y$ on $\RR^{1,5}$ with the same chirality obey
\begin{equation}\label{eq:6dselfdualidentity}
X_{\mu\nu\tau} Y^{\rho\sigma\tau} + Y_{\mu\nu\tau} X^{\rho\sigma\tau} = 2 \, \delta_{[\mu}^{[\rho} X^{\sigma ]\alpha\beta} Y_{\nu ] \alpha\beta}~.
\end{equation}
Furthermore $X_{\mu\rho\sigma} Y^{\nu\rho\sigma} = Y_{\mu\rho\sigma} X^{\nu\rho\sigma}$ and $X_{\mu\nu\rho} Y^{\mu\nu\rho} =0$ which imply the useful subsidiary identities 
\begin{equation}\label{eq:6dselfdualidentity2}
X_{\mu\rho\sigma} Y_{\nu}^{\;\, \rho\sigma} \, \Gamma^\mu \Gamma^\nu = 0 \; , \quad\quad X_{\mu\nu\tau} Y_{\rho\sigma}^{\;\;\; \tau} \, \Gamma^{\mu\nu\rho} \Gamma^\sigma = 0~.
\end{equation}
A useful corollary which follows using \eqref{eq:6dgammadual} is that ${\slashed X} \psi_\pm = 0$ identically for any symplectic Majorana-Weyl spinor $\psi_\pm$ and $X \in \bigwedge^3_\pm \RR^{1,5}$.

For any non-zero bosonic symplectic Majorana-Weyl spinor $\epsilon$ with positive chirality, only the bilinears ${\overline \epsilon}^A \Gamma_\mu \epsilon^B$ and $\zeta_{\mu\nu\rho}^{AB} = {\overline \epsilon}^A \Gamma_{\mu\nu\rho} \epsilon^B$ do not vanish identically.  Furthermore \eqref{eq:6dbilnears} implies ${\overline \epsilon}^A \Gamma_\mu \epsilon^B = - {\overline \epsilon}^B \Gamma_\mu \epsilon^A = \half \varepsilon^{AB} \xi_\mu$ and $\zeta_{\mu\nu\rho}^{AB} = \zeta_{\mu\nu\rho}^{BA}$, where the Dirac current $\xi_\mu = {\overline \epsilon} \Gamma_\mu \epsilon$ is the only non-zero contracted bilinear. Each three-form $\zeta^{AB}$ is self-dual since $\epsilon$ has positive chirality. 

The Fierz identity \eqref{eq:6dfierz} for $\epsilon$ reads
\begin{equation}\label{eq:6dfierzepsilon}
\epsilon^A {\overline \epsilon}^B = -\tfrac{1}{8} ( \varepsilon^{AB} {\slashed \xi} + {\slashed \zeta}^{AB} ) {\bf P}_-~,
\end{equation}
and a useful subsidiary identity is
\begin{equation}\label{eq:6dfierzepsilon2}
- \Gamma^{\mu\nu} \epsilon^A {\overline \epsilon}^B \, \Gamma_\nu = \left[ \left(  \epsilon^A {\overline \epsilon}^B + \half \varepsilon^{AB} {\slashed\xi} \right) \Gamma^\mu - \varepsilon^{AB} \xi^\mu {\bf 1} \right] {\bf P}_+~.
\end{equation}

Each self-dual $3$-form $\zeta^{AB} = {*\zeta}^{AB}$ obeys identically $\xi^\mu \zeta_{\mu\nu\rho}^{AB} =0$, implying that $\zeta_{\mu\nu\rho}^{AB} = 3\, \xi_{[\mu} \Omega_{\nu\rho ]}^{AB}$ in terms of a triple of $2$-forms $\Omega^{AB}$ on $\RR^{1,5}$ obeying $\xi^\mu \Omega_{\mu\nu}^{AB} =0$. 

The isotropy algebra $\fh_\epsilon$ of $\epsilon$ is isomorphic to $\fsu(2) \ltimes \RR^4 < \fso(1,5)$ and $\fk_\epsilon \cong ( \fsu(2) \ltimes \RR^4 ) \oplus \RR$. Relative to a null orthonormal basis $( {\bf e}_+ , {\bf e}_- , {\bf e}_a )$ on $\RR^{1,5}$, with $\eta_{+-} = 1$ and $\eta_{ab} = \delta_{ab}$ on $\RR^4 \subset \RR^{1,5}$, let us fix $\xi = {\bf e}_+$ so that $\Gamma_+ \epsilon^A =0$. The orientation tensor on $\RR^4$ is $\varepsilon_{abcd}$ with $\varepsilon_{1234} = \varepsilon_{+-1234} = 1$ and $\Gamma_{ab} \epsilon^A = - \half \varepsilon_{abcd} \Gamma^{cd} \epsilon^A$. The components $\Omega^{AB}_{ab} = {\overline \epsilon}^A \Gamma_{ab} \Gamma_- \epsilon^B$ define a quaternionic structure on $\RR^4$ in terms of a triple of anti-self-dual two-forms. They also provide an isomorphism between $\fusp(2)$ and $\bigwedge^2_- \RR^4$, as three-dimensional vector spaces. In this regard, a rather useful identity is $\varepsilon^{A(B} \epsilon^{C)} = \tfrac{1}{4} \Omega^{BC}_{ab} \Gamma^{ab} \epsilon^A$. Adapting this basis to a local frame on $\eM$ allows us to express 
\begin{equation}\label{eq:6dtorsion}
\nabla_\mu \epsilon^A = a_\mu \epsilon^A + b_\mu^a \Gamma_a \Gamma_-\epsilon^A + \half \, c_\mu^{ab} \Gamma_{ab} \epsilon^A~,
\end{equation}
in terms of real one-forms $a$, $b$ and $c$ valued respectively in the $\RR$, $\RR^4$ and $\fsu(2)$ factors of $\fk_\epsilon$, corresponding to the intrinsic torsion associated with the $\fsu(2) \ltimes \RR^4$-structure. (Additional terms of the form $\alpha^{AB}_\mu \epsilon_B + \beta^{AB}_{\mu a} \Gamma^a \Gamma_-\epsilon_B + \half \, \gamma^{AB}_{\mu ab} \Gamma^{ab} \epsilon_B$ which one might also have anticipated on the right hand side of \eqref{eq:6dtorsion} can be absorbed by a redefinition of $a$, $b$ and $c$ via the identity $\varepsilon^{A(B} \epsilon^{C)} = \tfrac{1}{4} \Omega^{BC}_{ab} \Gamma^{ab} \epsilon^A$.) Notice that $\nabla_\mu \xi_\nu$ does not involve $c$. Demanding that $\xi$ be a conformal Killing vector fixes $a_- =0$, $a^a = - b_-^a$, $b_+^a =0$ and $b_{ab} + b_{ba} = \delta_{ab} a_+$. Consequently, from \eqref{eq:6dtorsion}, one derives the twistor spinor equation $\eD_\mu \epsilon = \tfrac{1}{6} \Gamma_\mu {\slashed \eD} \epsilon$ with respect to connection $\eD_\mu = \nabla_\mu + {\sf t}_\mu$, where ${\sf t}_\mu = B_{\mu a} \, \Gamma^a \Gamma_- + \half \, C_\mu^{ab} \Gamma_{ab}$. The component $B_{\mu a}$ has $B_{+a} = 0 = B_{-a}$ and $B_{ba} = - \half b_{[ba]} - \tfrac{1}{4} \varepsilon_{bacd} b^{cd}$, whence spanning $\fsu(2) < \fh_\epsilon$. The other component $C_\mu^{ab}$ has $C_+^{ab} = - c_+^{ab} + \half b^{[ab]} - \tfrac{1}{4} \varepsilon^{abcd} b_{cd}$, $C_-^{ab} = - c_-^{ab}$, $C_c^{ab} = - c_c^{ab} + \delta_c^{[a} a^{b]} - \half \varepsilon_{c}{}^{abd} a_d$, whence spanning $T^* \eM \otimes \fsu(2)$ with the factor $\fsu(2) < \fk_\epsilon$. 

Using the identity $\varepsilon^{A(B} \epsilon^{C)} = \tfrac{1}{4} \Omega^{BC}_{ab} \Gamma^{ab} \epsilon^A$, it will be more convenient henceforth to express the $\half C_\mu^{ab} \Gamma_{ab} \epsilon^A$ term in $\eD_\mu \epsilon^A$ as $ C_\mu^A{}_B \, \epsilon^B$, where $C_\mu^A{}_B$ is a $\fusp(2)$-valued one-form on $\eM$, i.e. $( C_\mu^A{}_B )^* = - C_\mu^B{}_A$ and $C_\mu^A{}_A =0$. In the absence of the $B_{\mu a}$ component, integrability of the twistor spinor equation $\eD_\mu \epsilon = \tfrac{1}{6} \Gamma_\mu {\slashed \eD} \epsilon$ implies
\begin{equation}\label{eq:6dtwistorint}
\tfrac{1}{4} W_{\mu\nu\rho\sigma} \Gamma^{\rho\sigma} \epsilon^A = - \tfrac{3}{5} G_{\mu\nu}^A{}_B \epsilon^B + \tfrac{3}{20} G_{\mu\rho}^A{}_B \Gamma_{\nu}{}^\rho \epsilon^B - \tfrac{3}{20} G_{\nu\rho}^A{}_B \Gamma_{\mu}{}^\rho \epsilon^B - \tfrac{1}{40} \varepsilon_{\mu\nu\rho\sigma\alpha\beta} \, G^{\rho\sigma \, A}{}_B \Gamma^{\alpha\beta} \epsilon^B~,
\end{equation}
where $G^A{}_B = d C^A{}_B + C^A{}_C \wedge C^C{}_B$ is the curvature of $C^A{}_B$. In this case, a background $\eM$ is locally maximally supersymmetric only if $G^A{}_B =0$ with vanishing Weyl tensor. 

From equation (2.26) in \cite{Bergshoeff:1985mz}, one finds that $\eD$ can be identified with a special case of the superconnection for bosonic supersymmetric backgrounds of minimal conformal supergravity in six dimensions by relating ${\sf t}$ with components of the R-symmetry gauge field and self-dual three-form flux (written respectively as $V_\mu^{ij}$ and $T^-_{abc}$ in \cite{Bergshoeff:1985mz}). However, any such identification must also be compatible with the additional constraints which come from setting to zero the supersymmetry variation of the dilatino in the conformal supergravity background. The simple case of a twistor spinor with respect to $\eD = \nabla + C$ corresponds to a bosonic conformal supergravity background with no three-form flux. In that case, the additional constraint from the dilatino variation fixes the background value of the dilaton $\varphi$ such that $\varphi \, \epsilon^A$ is a particular linear combination of $G_{\mu\nu}^A{}_B \Gamma^{\mu\nu} \epsilon^B$ and $( \nabla^\mu C_\mu^A{}_B ) \epsilon^B$.  

\subsection{Yang-Mills supermultiplet}
\label{sec:6dSYMonshell} 

The minimal on-shell Yang-Mills supermultiplet in ${\sf d}=6$ contains a bosonic gauge field $A_\mu$ and a fermionic symplectic Majorana-Weyl spinor $\lambda^A$ (we take $\lambda^A$ with positive chirality, i.e. $\Gamma \lambda^A = \lambda^A$). To match the $8$ off-shell fermionic degrees of freedom of $\lambda^A$, the $5$ off-shell degrees of freedom of the gauge field can be supplemented by $3$ real bosonic auxiliary scalars, in the form of a complex triple $Y^{AB} = Y^{BA}$ subject to reality condition $( Y^{AB} )^* = \varepsilon_{AC} \varepsilon_{BD} Y^{CD}$ (i.e. in the adjoint representation of the auxiliary $\fusp(2)$). 

The supersymmetry transformations on $\RR^{1,5}$ are     
\begin{align}\label{eq:6dsusy}
\delta_\epsilon A_\mu &= {\overline \epsilon}^A \Gamma_\mu \lambda_A \nonumber \\
\delta_\epsilon \lambda^A &= -\tfrac{1}{2} F^{\mu\nu} \Gamma_{\mu\nu} \epsilon^A + Y^{AB} \epsilon_B \nonumber \\
\delta_\epsilon Y^{AB} &= {\overline \epsilon}^A {\slashed D} \lambda^B + {\overline \epsilon}^B {\slashed D} \lambda^A~,
\end{align}
where $\epsilon$ is a constant bosonic symplectic Majorana-Weyl spinor with positive chirality. Squaring \eqref{eq:6dsusy} gives $\delta_\epsilon^2 = \xi^\mu \partial_\mu + \delta_\Lambda$ identically with gauge parameter $\Lambda = -\xi^\mu A_\mu$ and therefore the supersymmetry algebra closes off-shell. 

Up to boundary terms, the lagrangian 
\begin{equation}\label{eq:6dsusylagoffshell}
\eL_{\mathrm{SYM}} = -\tfrac{1}{4} ( F_{\mu\nu} , F^{\mu\nu} ) -\half ( {\overline \lambda}^A , {\slashed D} \lambda_A ) + \tfrac{1}{4} ( Y^{AB} , Y_{AB} )~,
\end{equation}
is invariant under \eqref{eq:6dsusy}. It is not scale-invariant but has a $\fusp(2)$ R-symmetry, under which $A_\mu$ is a singlet, $\lambda^A$ is in the fundamental and $Y^{AB}$ is in the adjoint representation. In particular, the R-symmetry variations are $\delta_\rho A_\mu = 0$, $\delta_\rho \lambda^A = \rho^{A}_{\;\;\;\; B} \lambda^B$ and $\delta_\rho Y^{AB} = \rho^{A}_{\;\;\;\; C} Y^{BC} + \rho^{B}_{\;\;\;\; C} Y^{AC}$, with parameter obeying $\rho^{A}_{\;\;\;\; B} = - ( \rho^{B}_{\;\;\;\; A} )^*$ and $\rho^{A}_{\;\;\;\; A} =0$ in the adjoint of $\fusp(2)$. 

We now consider the minimally coupled version of \eqref{eq:6dsusy} on a six-dimensional lorentzian spin manifold $\eM$. Squaring them gives  
\begin{align}\label{eq:6dsusysquared}
\delta_\epsilon^2 A_\mu &= \cL_\xi A_\mu + D_\mu \Lambda \nonumber \\
\delta_\epsilon^2 \lambda^A &= \xi^\mu \nabla_\mu \lambda^A + [ \lambda^A , \Lambda ] - \Gamma^{\mu\nu} \epsilon^A ( \nabla_\mu {\overline \epsilon}^B ) \Gamma_\nu \lambda_B \\
\delta_\epsilon^2  Y^{AB} &= \cL_\xi  Y^{AB} +  [ Y^{AB} , \Lambda ] + ( {\overline \epsilon}^A {\slashed \nabla} \epsilon_C ) Y^{BC} + ( {\overline \epsilon}^B {\slashed \nabla} \epsilon_C ) Y^{AC} - \half F_{\mu\nu} \left( {\overline \epsilon}^A \Gamma^\rho \Gamma^{\mu\nu} \nabla_\rho \epsilon^B + {\overline \epsilon}^B \Gamma^\rho \Gamma^{\mu\nu} \nabla_\rho \epsilon^A \right)\nonumber~.
\end{align}
If $\xi$ is a Killing vector, off-shell closure of the supersymmetry algebra requires $\delta_\epsilon^2 = \cL_\xi + \delta_\Lambda$. This is tantamount to imposing
\begin{equation}\label{eq:6dsusyclosure}
{\overline \epsilon}^A {\slashed \nabla} \epsilon^B =0 \; , \quad\quad {\overline \epsilon}^A \Gamma_{[\mu} \nabla_{\nu ]} \epsilon^B + {\overline \epsilon}^B \Gamma_{[\mu} \nabla_{\nu ]} \epsilon^A = 0 \; , \quad\quad \nabla_{[\mu} \xi_{\nu ]} =  {\overline \epsilon}^A \Gamma_{\mu\nu\rho} \nabla^\rho \epsilon_A \; , \quad\quad \nabla^\rho \zeta_{\mu\nu\rho}^{AB} = 0~.
\end{equation}

They are solved if $\epsilon$ satisfies
\begin{equation}\label{eq:6dTspinor}
\nabla_\mu \epsilon^A = \tfrac{1}{8} h_{\mu\nu\rho} \Gamma^{\nu\rho} \epsilon^A~,
\end{equation}
for some self-dual three-form $h$. Any such $\epsilon$ is necessarily harmonic (i.e. ${\slashed \nabla} \epsilon^A =0$) following the corollary noted below \eqref{eq:6dselfdualidentity2}. 

Furthermore, up to boundary terms, the lagrangian 
\begin{equation}\label{eq:6dsusylagoffshellH}
-\tfrac{1}{4} ( F_{\mu\nu} , F^{\mu\nu} ) -\half ( {\overline \lambda}^A , {\slashed D} \lambda_A ) + \tfrac{1}{4} ( Y^{AB} , Y_{AB} ) + \half\, h^{\mu\nu\rho} ( A_\mu , \partial_\nu A_\rho + \tfrac{1}{3} [ A_\nu , A_\rho ] ) ~,
\end{equation}
on $\eM$ is invariant under the minimal coupling of \eqref{eq:6dsusy} provided $h$ is closed. 

Unlike the minimal coupling of Yang-Mills supersymmetry transformations in lower dimensions, notice that the minimal coupling of \eqref{eq:6dsusy} is not automatically Weyl-covariant. However, the supersymmetry transformations can be conformally coupled by adding an improvement term $\tfrac{1}{3} ( {\overline \lambda}^A {\slashed \nabla} \epsilon^B  +  {\overline \lambda}^B {\slashed \nabla} \epsilon^A )$ to the right hand side of $\delta_\epsilon Y^{AB}$, relative to the minimal coupling of \eqref{eq:6dsusy} on $\eM$. If $\epsilon$ is a twistor spinor, one finds that $\delta_\epsilon^2 = \cL_\xi + \delta_\sigma + \delta_\Lambda + \delta_\rho$ off-shell, where $\delta_\sigma$ is a Weyl variation with parameter $\sigma = -\tfrac{1}{6} \nabla_\mu \xi^\mu$ and $\delta_\rho$ is a $\fusp(2)$ R-symmetry variation with parameter $\rho^{AB} = \tfrac{2}{3} ( {\overline \epsilon}^A {\slashed \nabla} \epsilon^B  +  {\overline \epsilon}^B {\slashed \nabla} \epsilon^A )$. The fields $( A_\mu , \lambda^A , Y^{AB} )$ are assigned their canonical Weyl weights $(0,-\tfrac{3}{2},-2)$.

If $\epsilon$ is a twistor spinor with respect to the connection $\eD = \nabla + C$, the same picture emerges after gauging the $\fusp(2)$ R-symmetry. This means that a local R-symmetry variation $\delta_\rho \Phi^A = \rho^A{}_B \Phi^B$ of a field $\Phi^A$ in the fundamental representation of $\fusp(2)$ is accompanied by the variation $\delta_\rho C^A{}_B = - d \rho^A{}_B$. Therefore $\eD \Phi^A = d \Phi^A + C^A{}_B \, \Phi^B$ transforms covariantly under $\delta_\rho$. The R-symmetry gauging is tantamount to replacing $\nabla$ with $\eD$ in the Weyl-covariant expressions above. One then takes the $\fusp(2)$ gauge field $C^A{}_B$ to be a fixed background one-form with Weyl weight zero.

The integral of the minimally coupled Yang-Mills lagrangian \eqref{eq:6dsusylagoffshell} on $\eM$ is not Weyl-invariant and cannot be conformally coupled without additional compensator fields. If $\epsilon$ is a twistor spinor, up to boundary terms, one finds $\delta_\epsilon \eL_{\mathrm{SYM}} = \tfrac{1}{6} \, ( F^{\mu\nu} , {\overline \lambda}_A \Gamma_{\mu\nu} {\slashed \nabla} \epsilon^A ) + \tfrac{1}{3} \, ( Y_{AB} , {\overline \lambda}^A {\slashed \nabla} \epsilon^B )$ so the minimally coupled lagrangian is generically supersymmetric only if $\nabla_\mu \epsilon =0$.

However, now let
\begin{equation}\label{eq:6dCKS}
\nabla_\mu \epsilon^A = \alpha_\nu \Gamma_\mu \Gamma^\nu \epsilon^A -\tfrac{1}{8}\, \beta_{\mu\nu\rho} \Gamma^{\nu\rho} \epsilon^A~,
\end{equation}
where $\alpha$ is a real one-form and $\beta$ is a real anti-self-dual three-form. Integrability of this algebraic twistor spinor equation implies $\alpha$ must be closed. (Note that additional terms of the form $\gamma_\nu^{AB} \Gamma_\mu \Gamma^\nu \epsilon_B -\tfrac{1}{8}\, \delta_{\mu\nu\rho}^{AB} \Gamma^{\nu\rho} \epsilon_B$ on the right hand side of \eqref{eq:6dCKS} can always be absorbed by a redefinition of $\alpha$ and $\beta$ using the identity $\varepsilon^{A(B} \epsilon^{C)} = \tfrac{1}{4} \Omega^{BC}_{ab} \Gamma^{ab} \epsilon^A$.) 

If $\alpha = \half d \varphi$ in \eqref{eq:6dCKS}, for some function $\varphi$, one finds that the lagrangian 
\begin{equation}\label{eq:6dsusylagoffshelltwistorcorrection}
{\mbox{e}}^{-2\, \varphi} \left[ -\tfrac{1}{4} ( F_{\mu\nu} , F^{\mu\nu} ) -\half ( {\overline \lambda}^A , {\slashed D} \lambda_A ) + \tfrac{1}{4} ( Y^{AB} , Y_{AB} ) + \half\, \beta^{\mu\nu\rho} ( A_\mu , \partial_\nu A_\rho + \tfrac{1}{3} [ A_\nu , A_\rho ] ) + \tfrac{1}{48} \, \beta^{\mu\nu\rho}  ( {\overline \lambda}^A ,\Gamma_{\mu\nu\rho} \lambda_A ) \right]~,
\end{equation}
is supersymmetric provided $d ( {\mbox{e}}^{-2\, \varphi} \beta ) =0$. The prefactor ${\mbox{e}}^{-2\, \varphi}$ in \eqref{eq:6dsusylagoffshelltwistorcorrection} plays the r\^{o}le of a conformal compensator while $\beta$ couples to a Chern-Simons term for the gauge field and a mass term for the gaugino. One can use a Weyl transformation to fix $\varphi =0$. In this case $\epsilon$ obeys $\nabla_\mu \epsilon^A = -\tfrac{1}{8} \beta_{\mu\nu\rho} \Gamma^{\nu\rho} \epsilon^A$ with $\beta$ closed and anti-self-dual. Notice the form of the corrections here (in terms of $\beta$) are almost identical to those from \eqref{eq:6dTspinor} and \eqref{eq:6dsusylagoffshellH} (in terms of $h$). The fact that $h$ is self-dual whereas $\beta$ is anti-self-dual is rather significant though since it is only the latter which is compatible with the conformal structure and necessitates the non-vanishing fermionic mass term in \eqref{eq:6dsusylagoffshelltwistorcorrection}.

If $\epsilon$ is an algebraic twistor spinor of the form \eqref{eq:6dCKS} but with respect to $\eD = \nabla + C$, the same story follows after gauging the R-symmetry and replacing $\nabla$ with $\eD$ in the expressions above.

\subsubsection{Supergravity backgrounds}
\label{sec:6dsugrabackgrounds} 

Solutions of \eqref{eq:6dTspinor} with $dh=0$ correspond to bosonic supersymmetric backgrounds of minimal Poincar\'{e} supergravity in six dimensions. A classification of all such backgrounds which solve the supergravity equations of motion was obtained in \cite{Gutowski:2003rg} and also \cite{Chamseddine:2003yy} for the case of maximally supersymmetric solutions. 

Symplectic Majorana-Weyl spinors in ${\sf d}=6$ can be thought of locally as vectors in $\HH^{2}$ so there can be no more than two linearly independent quaternionic $\epsilon$ which solve \eqref{eq:6dTspinor}. Whence any bosonic supersymmetric background is either maximally or minimally supersymmetric. We defer to \cite{Gutowski:2003rg} for a description of the minimally supersymmetric solutions. 

As shown in \cite{Chamseddine:2003yy}, there is a one-to-one correspondence between the maximally supersymmetric backgrounds and isomorphism classes of six-dimensional lie groups equipped with bi-invariant lorentzian metric and self-dual parallelising torsion. This restricts all maximally supersymmetric geometries to be locally isometric to either $\RR^{1,5}$, $\AdS_3 \times S^3$ or a plane wave. Any such solution admits a trivial oxidation to a half-BPS solution within the first class of supergravity vacua in ten dimensions to be discussed in section~\ref{sec:sugrabackgrounds}. 

The algebraic twistor spinor equation $\nabla_\mu \epsilon^A = -\tfrac{1}{8} \beta_{\mu\nu\rho} \Gamma^{\nu\rho} \epsilon^A$ with $\beta$ closed and anti-self-dual has a simple solution on $\eM = \AdS_3 \times S^3$ with $\beta = \half \gamma\, ( {\mbox{vol}}_{\AdS_3} -  {\mbox{vol}}_{S^3} )$, for any real constant $\gamma$. The twistor spinor $\epsilon$ on $\eM$ decomposes into a tensor product of Killing spinors on the $\AdS_3$ and $S^3$ factors (with Killing constant $\gamma$ on $\AdS_3$ and $i\gamma$ on $S^3$). Indeed, it is related to the maximally supersymmetric Freund-Rubin $\AdS_3 \times S^3$ solution of minimal Poincar\'{e} supergravity mentioned above by a sign change in the relative orientation between the $\AdS_3$ and $S^3$ factors.

\subsection{Matter supermultiplet}
\label{sec:6dmatteronshell} 

The minimal on-shell matter supermultiplet in ${\sf d}=6$ contains four real bosonic scalars $\phi^{A{\dot B}}$ and a fermionic Weyl spinor $\psi^{\dot A}$ (we take $\psi^{\dot A}$ with negative chirality, i.e. $\Gamma \psi^{\dot A} = -\psi^{\dot A}$). Dotted indices denote the complex doublet representation of a $\fusp(2)$ global symmetry for the matter supermultiplet. This is distinct from the $\fusp(2)$ R-symmetry acting on undotted indices encountered above (though we use the same tensorial conventions for both $\fusp(2)$ factors). The four bosonic scalars obey the reality condition $( \phi^{A{\dot B}} )^* = \varepsilon_{AC} \,\varepsilon_{{\dot B}{\dot D}} \phi^{C{\dot D}}$, defining the vector representation of $\fso(4) \cong \fusp(2) \oplus \fusp(2)$. The fermionic Weyl spinors obey $( \psi^{\dot A} )^* = \varepsilon_{{\dot A}{\dot B}} \, {\bf B} \psi^{\dot B}$, a symplectic Majorana condition but with respect to the other $\fusp(2)$ factor. The $4$ on-shell fermionic degrees of freedom of $\psi^{\dot A}$ match the $4$ on-shell bosonic degrees of freedom of $\phi^{A{\dot B}}$. Both fields are valued in a (real form of a) quaternionic unitary representation $V$ of $\fg$. The $\fg$-invariant inner product $\langle -,- \rangle$ is a real symmetric $\CC$-bilinear form on $V$.

The supersymmetry transformations for the matter supermultiplet coupled to the minimal gauge supermultiplet on $\RR^{1,5}$ are     
\begin{align}\label{eq:6dsusymatter}
\delta_\epsilon \phi^{A{\dot B}} &= {\overline \epsilon}^A \psi^{\dot B} \nonumber \\
\delta_\epsilon \psi^{\dot A} &= 2 \, ( {\slashed D} \phi^{B{\dot A}} ) \, \epsilon_B~,
\end{align}
where $\epsilon^A$ is a constant bosonic symplectic Majorana-Weyl spinor with positive chirality. Squaring \eqref{eq:6dsusymatter} gives $\delta_\epsilon^2 = \xi^\mu \partial_\mu + \delta_\Lambda$ after imposing fermionic equation of motion ${\slashed D} \psi^{\dot A} + 2 \, \lambda_B \cdot \phi^{B{\dot A}} =0$. Whence, the supersymmetry algebra closes on-shell. 

Up to boundary terms, the lagrangian 
\begin{equation}\label{eq:6dsusymatterlag}
\eL_{\mathrm{M}} = -\half  \langle D_\mu \phi^{A{\dot B}} , D^\mu \phi_{A{\dot B}} \rangle  -\tfrac{1}{4} \langle {\overline \psi}^{\dot A} , {\slashed D} \psi_{\dot A} \rangle + \half  \langle \phi_{A{\dot B}} , Y^{A}_{\;\;\;\;\; C} \cdot \phi^{C{\dot B}} \rangle + \langle {\overline \psi}_{\dot A} , \lambda_B \cdot \phi^{B{\dot A}}\rangle~,
\end{equation}
is invariant under \eqref{eq:6dsusy} and \eqref{eq:6dsusymatter}. The integral of $\eL_{\mathrm{M}}$ on $\RR^{1,5}$ is classically scale-invariant, with $( \phi^{A{\dot B}} , \psi^{\dot A} )$ having dimensions $(2,\tfrac{5}{2})$ and $( A_\mu , \lambda^A , Y^{AB} )$ assigned their canonical dimensions $(1,\tfrac{3}{2}, 2)$. It is also manifestly invariant under the $\fusp(2) \oplus \fusp(2)$ global symmetry.

The conformal coupling of \eqref{eq:6dsusymatter} and \eqref{eq:6dsusymatterlag} on $\eM$ follows via the improvement terms described in \eqref{eq:conformallycoupledsusy} and \eqref{eq:conformallycoupledscalar}. Relative to their minimal coupling on $\eM$, this involves adding a term $\tfrac{4}{3} \phi^{B{\dot A}} {\slashed \nabla} \epsilon_B$ to the right hand side of $\delta_\epsilon \psi^{\dot A}$ in \eqref{eq:6dsusymatter} and a term $- \tfrac{1}{10} \, R \langle \phi^{A{\dot B}} , \phi_{A{\dot B}} \rangle$ to $\eL_{\mathrm{M}}$. If $\epsilon$ is a twistor spinor on $\eM$, the conformally coupled supersymmetry transformations for the matter fields close on-shell with $\delta_\epsilon^2 = \cL_\xi + \delta_\sigma +\delta_\Lambda + \delta_\rho$, where $( \phi^{A{\dot B}} , \psi^{\dot A} )$ are assigned their canonical weights $(-2,-\tfrac{5}{2})$, with Weyl parameter $\sigma = -\tfrac{1}{6} \nabla_\mu \xi^\mu$ and R-symmetry parameter $\rho^{AB} = \tfrac{2}{3} ( {\overline \epsilon}^A {\slashed \nabla} \epsilon^B  +  {\overline \epsilon}^B {\slashed \nabla} \epsilon^A )$, just as for the conformally coupled off-shell Yang-Mills supermultiplet. Furthermore, the integral of the conformally coupled matter lagrangian on $\eM$ is invariant under these supersymmetry transformations. 

If $\epsilon$ is a twistor spinor with respect to $\eD = \nabla + C$, the same story follows after gauging the R-symmetry and replacing $\nabla$ with $\eD$ in the expressions above.

\subsection{Tensor supermultiplet}
\label{sec:6dtensoronshell} 

The minimal on-shell abelian tensor supermultiplet in ${\sf d}=6$ contains a bosonic $2$-form gauge field $B_{\mu\nu}$, a real bosonic scalar $\phi$ and a fermionic symplectic Majorana-Weyl spinor $\chi^A$. We take the field strength $H =dB$ to be anti-self-dual ($H = -{*H}$), $\chi^A$ with negative chirality ($\Gamma \chi^A = -\chi^A$) and ${\slashed \partial} \chi^A =0$ as its equation of motion. The $4$ on-shell fermionic degrees of freedom of $\chi^A$ are matched by the $3$ on-shell bosonic degrees of freedom of $B_{\mu\nu}$ plus one of $\phi$. All fields are abelian. 

The supersymmetry transformations on $\RR^{1,5}$ are     
\begin{align}\label{eq:6dsusytensor}
\delta_\epsilon B_{\mu\nu} &= {\overline \epsilon}^A \Gamma_{\mu\nu} \chi_A \nonumber \\
\delta_\epsilon \chi^A &= -\tfrac{1}{12} \, H^{\mu\nu\rho} \Gamma_{\mu\nu\rho} \epsilon^A + \partial^\mu \phi \, \Gamma_\mu \epsilon^A \nonumber \\
\delta_\epsilon \phi &= {\overline \epsilon}^A \chi_A~,
\end{align}
where $\epsilon^A$ is a constant bosonic symplectic Majorana-Weyl spinor with positive chirality. The transformations in \eqref{eq:6dsusytensor} are scale-covariant; $\epsilon^A$ has dimension $-\half$ and $( B_{\mu\nu} , \chi^A , \phi )$ have must have dimensions $( \Delta , \Delta + \half , \Delta )$, for some constant $\Delta$. Squaring \eqref{eq:6dsusytensor} gives $\delta_\epsilon^2 = \xi^\mu \partial_\mu + \delta_\Omega$, after imposing equations of motion $H = -{*H}$ and ${\slashed \partial} \chi^A =0$, where $\delta_\Omega B = d\Omega$, $\delta_\Omega \chi^A = 0$ and $\delta_\Omega \phi =0$ under the abelian gauge symmetry with $1$-form gauge parameter $\Omega_\mu = B_{\mu\nu} \, \xi^\nu - \phi \, \xi_\mu$. Whence, the supersymmetry algebra closes on-shell. (Anti-)self-duality of $H$ obstructs the straightforward construction of a supersymmetric lagrangian for this theory.  

The conformal coupling of \eqref{eq:6dsusytensor} on $\eM$ follows via the improvement term described in \eqref{eq:conformallycoupledsusy}. Relative to the minimal coupling on $\eM$, this means adding a term $\tfrac{2}{3}\, \phi {\slashed \nabla} \epsilon^A$ to the right hand side of $\delta_\epsilon \chi^A$ in \eqref{eq:6dsusytensor}. The minimally coupled equations of motion $H = -{*H}$ and ${\slashed \nabla} \chi^A =0$ are automatically Weyl-invariant on $\eM$, provided $\chi^A$ has weight $-\tfrac{5}{2}$. If $\epsilon$ is a twistor spinor on $\eM$, the conformally coupled supersymmetry transformations close on-shell with $\delta_\epsilon^2 = \cL_\xi + \delta_\sigma +\delta_\Omega + \delta_\rho$, where $( B_{\mu\nu} , \chi^A , \phi )$ are assigned their canonical weights $(0,-\tfrac{5}{2},-2)$, the Weyl parameter $\sigma = -\tfrac{1}{6} \nabla_\mu \xi^\mu$ and the R-symmetry parameter $\rho^{AB} = \tfrac{2}{3} ( {\overline \epsilon}^A {\slashed \nabla} \epsilon^B  +  {\overline \epsilon}^B {\slashed \nabla} \epsilon^A )$.

A novel minimally supersymmetric coupling of the tensor multiplet to the off-shell Yang-Mills multiplet on $\RR^{1,5}$ was obtained in \cite{Bergshoeff:1996qm}. The supersymmetry transformations are
\begin{align}\label{eq:6dsusytensorvector}
\delta_\epsilon B_{\mu\nu} &= {\overline \epsilon}^A \Gamma_{\mu\nu} \chi_A + \kappa \, ( A_{\mu} , \delta_\epsilon A_{\nu } ) - \kappa \, ( A_{\nu} , \delta_\epsilon A_{\mu } ) \nonumber \\
\delta_\epsilon \chi^A &= -\tfrac{1}{12} \, \eH^{\mu\nu\rho} \Gamma_{\mu\nu\rho} \epsilon^A + \partial^\mu \phi \, \Gamma_\mu \epsilon^A + \tfrac{\kappa}{2} \, ( \delta_\epsilon A_\mu \, , \Gamma^\mu \lambda^A ) \nonumber \\
\delta_\epsilon \phi &= {\overline \epsilon}^A \chi_A ~,
\end{align}
where $\kappa$ is a constant and $\eH_{\mu\nu\rho} =  H_{\mu\nu\rho} + 6 \kappa\, ( A_{[\mu} , \partial_{\nu} A_{\rho ]} ) + 2\kappa\, ( A_{\mu} , [ A_\nu , A_\rho ] )$. In addition to the abelian tensor gauge symmetry under $\delta_\Omega B = d \Omega$, the field strength $\eH$ is also invariant under the infinitesimal gauge transformations $\delta_\Lambda B_{\mu\nu} = - 2\kappa\, ( \Lambda , \partial_{[ \mu} , A_{\nu ]} )$ and $\delta_\Lambda A_\mu = D_\mu \Lambda$, with $\delta_\Lambda \chi^A =0$ and $\delta_\Lambda \phi =0$. Squaring \eqref{eq:6dsusytensorvector} gives $\delta_\epsilon^2 = \xi^\mu \partial_\mu + \delta_\Omega + \delta_{\Lambda}$, with the same abelian $\Omega$ and $\fg$-valued $\Lambda$ gauge parameters as above, after imposing the equations of motion 
\begin{equation}\label{eq:6dsusytensoreom}
\eH^+_{\mu\nu\rho} =  -\tfrac{\kappa}{4} \, ( {\overline \lambda}^A , \Gamma_{\mu\nu\rho} \lambda_A ) \; , \quad\quad {\slashed \partial} \chi^A = \kappa \left( \half ( F^{\mu\nu} , \Gamma_{\mu\nu} \lambda^A ) + ( Y^{AB} , \lambda_B ) \right)~.
\end{equation}
If $( B_{\mu\nu} , \chi^A , \phi )$ have dimensions $( \Delta , \Delta + \half , \Delta )$ and $( A_\mu , \lambda^A , Y^{AB} )$ are assigned their canonical dimensions $(1,\tfrac{3}{2},2)$, \eqref{eq:6dsusytensorvector} and \eqref{eq:6dsusytensoreom} are scale-invariant provided $\kappa$ has dimension $\Delta -2$. Whence, $\kappa$ is dimensionless for $\Delta =2$.

The conformal coupling on $\eM$ proceeds just as for the ungauged theory, requiring only the addition of $\tfrac{2}{3}\, \phi {\slashed \nabla} \epsilon^A$ to the right hand side of $\delta_\epsilon \chi^A$ relative to the minimal coupling of \eqref{eq:6dsusytensorvector}. The minimal coupling of the equations of motion \eqref{eq:6dsusytensoreom} are automatically Weyl-invariant on $\eM$, provided $\chi^A$ has weight $-\tfrac{5}{2}$. If $\epsilon$ is a twistor spinor on $\eM$, the conformally coupled supersymmetry transformations close on-shell with $\delta_\epsilon^2 = \cL_\xi + \delta_\sigma +\delta_\Omega + \delta_\Lambda + \delta_\rho$, after imposing the minimally coupled field equations \eqref{eq:6dsusytensoreom}. 

If $\epsilon$ is a twistor spinor with respect to $\eD = \nabla + C$, the same story follows after gauging the R-symmetry and replacing $\nabla$ with $\eD$ in the expressions above.

Note that closure of the conformally coupled supersymmetry transformations for the tensor multiplet is off-shell with respect to the conformally coupled Yang-Mills supermultiplet. As we have already seen, the minimally coupled Yang-Mills lagrangian is not supersymmetric with respect to the conformally coupled transformations. Furthermore, the self-dual projection in the first equation in \eqref{eq:6dsusytensoreom} continues to hamper the construction of a lagrangian for the tensor multiplet. However, let us consider the special case where the supersymmetry parameter $\epsilon$ is an algebraic twistor spinor obeying $\nabla_\mu \epsilon^A = -\tfrac{1}{8} \, \beta_{\mu\nu\rho} \Gamma^{\nu\rho} \epsilon^A$, with $\beta$ closed and anti-self-dual. In this case, recall that supersymmetry is restored with lagrangian \eqref{eq:6dsusylagoffshelltwistorcorrection} for the Yang-Mills multiplet (after using a Weyl transformation to fix the conformal compensator). Remarkably, the coefficients of $\beta$ in this correction term for the Yang-Mills lagrangian are actually proportional to the $\kappa$-dependent terms in the first equation \eqref{eq:6dsusytensoreom} for the tensor multiplet! Indeed, by dropping the assumption that $\beta$ is closed, and thinking of it as a lagrange multipler, one can obtain the first equation in \eqref{eq:6dsusytensoreom} as a constraint by adding the term $\tfrac{1}{12\kappa} \beta^{\mu\nu\rho} H_{\mu\nu\rho}$ from the tensor multiplet to the supersymmetric Yang-Mills lagrangian. The integral of this term on $\eM$ is not supersymmetric on its own unless $\beta$ is closed, which is the equation of motion for $B$. Without this constraint though, supersymmetry can be restored by adding a final term $- \tfrac{1}{10\kappa} R \phi$ to the lagrangian. The equation $R=0$ follows as an integrability condition from the algebraic twistor spinor equation when $\beta$ is closed. To summarise, up to boundary terms, the lagrangian 
\begin{equation}\label{eq:6dsusygaugetensorcurved}
\eL_{\mathrm{SYM}} + \tfrac{1}{12\kappa} \, \beta^{\mu\nu\rho} \left( \eH^+_{\mu\nu\rho} + \tfrac{\kappa}{4} \, ( {\overline \lambda}^A , \Gamma_{\mu\nu\rho} \lambda_A ) \right) -\tfrac{1}{10\kappa} \, R \phi~, 
\end{equation}
is invariant under the conformally coupled supersymmetry transformations for the tensor and Yang-Mills multiplets with algebraic twistor spinor parameter $\epsilon$ obeying $\nabla_\mu \epsilon^A = -\tfrac{1}{8} \, \beta_{\mu\nu\rho} \Gamma^{\nu\rho} \epsilon^A$ and $\beta$ anti-self-dual. Similarly, if $\epsilon$ is an algebraic twistor spinor of this type but with respect to $\eD = \nabla + C$, the same results follow after gauging the R-symmetry and replacing $\nabla$ with $\eD$ in the expressions above.

Note that the correction terms involving fields from the tensor multiplet vanish identically in Minkowski space. This rather odd way of deriving the field equation $\eH^+_{\mu\nu\rho} =- \tfrac{\kappa}{4} \, ( {\overline \lambda}^A , \Gamma_{\mu\nu\rho} \lambda_A )$ from \eqref{eq:6dsusygaugetensorcurved} is therefore only possible when $\eM$ is curved. The absence of terms involving $\chi^A$ in \eqref{eq:6dsusygaugetensorcurved} is also noteworthy. The supersymmetry variation of the minimal coupling of the second equation in \eqref{eq:6dsusytensoreom}, that is associated with the equation of motion for $\chi^A$, gives 
\begin{equation}\label{eq:6dsusychieom}
-\half \nabla^\rho [ \eH^+_{\mu\nu\rho} + \tfrac{\kappa}{4} \, ( {\overline \lambda}^B , \Gamma_{\mu\nu\rho} \lambda_B ) ] \Gamma^{\mu\nu} \epsilon^A + [ \nabla^2 \phi -\tfrac{1}{5} R \phi + 2\kappa \eL_{\mathrm{SYM}}] \epsilon^A~.
\end{equation}
The equation of motion for $\phi$ in \eqref{eq:6dsusychieom} actually follows from setting to zero the Weyl variation of the metric in \eqref{eq:6dsusygaugetensorcurved}. One might have expected terms proportional to
\begin{equation}\label{eq:6dsusyexpectedterms}
-\half \nabla_\mu \phi \nabla^\mu \phi -\tfrac{1}{10} R \phi^2 + 2\kappa  \phi \, \eL_{\mathrm{SYM}} + \tfrac{1}{2} \, {\overline \chi}_A {\slashed \nabla} \chi^A -  \tfrac{\kappa}{2} ( F^{\mu\nu} ,  {\overline \chi}_A \Gamma_{\mu\nu} \lambda^A ) - \kappa ( Y^{AB} , {\overline \chi}_A \lambda_B )~,
\end{equation} 
in the lagrangian, to yield the correct equations of motion for $\chi^A$ and $\phi$. The reason this is not necessary here is that their equations of motion can be generated by taking supersymmetry variations of $\eH^+_{\mu\nu\rho} =- \tfrac{\kappa}{4} \, ( {\overline \lambda}^A , \Gamma_{\mu\nu\rho} \lambda_A )$. In a supersymmetric field theory, one typically expects to obtain second order bosonic equations of motion by varying first order fermionic ones. The tensor supermultiplet is exceptional in that it contains a first order bosonic field equation that generates them all. 

\section{${\sf d}=10$}
\label{sec:10dSUSY} 

The clifford algebra $\Cl(1,9) \cong {\mbox{Mat}}_{32} ( \RR )$ and its action on $\RR^{32}$ defines the Majorana spinor representation. The $\pm 1$ eigenspaces of $\Gamma$ under the action of ${\mbox{Mat}}_{32} ( \RR )$ are isomorphic to $\RR^{16}$ and define the $\pm$ chirality Majorana-Weyl spinor representations.

The identities in \eqref{eq:GammaProducts} imply
\begin{equation}\label{eq:10dgammaidentities}
\Gamma^\alpha \Gamma_\mu \Gamma_\alpha = -8\, \Gamma_\mu \; , \quad\quad \Gamma^\alpha \Gamma_{\mu\nu\rho} \Gamma_\alpha = -4\, \Gamma_{\mu\nu\rho}  \; , \quad\quad \Gamma^\alpha \Gamma_{\mu\nu\rho\sigma\tau} \Gamma_\alpha = 0  ~,
\end{equation} 
and 
\begin{equation}\label{eq:10dgammadual}
\Gamma_{\mu_1 ... \mu_k} \Gamma = \sigma_{k-1} \, \tfrac{1}{(10-k)!} \, \varepsilon_{\mu_1 ... \mu_k \nu_{k+1} ... \nu_{10}} \Gamma^{\nu_{k+1} ... \nu_{10}}~,
\end{equation}
where $\varepsilon_{01...9} =1$ and $\Gamma = - \Gamma_0 \Gamma_1 ... \Gamma_9$.

The charge conjugation matrix obeys $( {\bf C} \Gamma_{\mu_1 ... \mu_k} )^t = - \sigma_k \, {\bf C} \Gamma_{\mu_1 ... \mu_k}$ and so ${\overline{\Gamma_{\mu_1 ... \mu_k} \epsilon}} = \sigma_k \, {\overline \epsilon} \Gamma_{\mu_1 ... \mu_k}$ for any Majorana spinor $\epsilon$. Thus any two fermionic Majorana spinors $\psi$ and $\chi$ obey
\begin{equation}\label{eq:10dbilnears}
{\overline \chi} \Gamma_{\mu_1 ... \mu_k} \psi = \sigma_k \, {\overline \psi} \Gamma_{\mu_1 ... \mu_k} \chi~,
\end{equation}
(For either $\psi$ or $\chi$ bosonic, one just puts an overall minus sign in \eqref{eq:10dbilnears}.)  

Any two Majorana-Weyl spinors $\psi_\pm$ and $\chi_\pm$ (with the same $\pm$ chirality) obey
\begin{equation}\label{eq:10devenbilnears}
{\overline \chi_\pm} \Gamma_{\mu_1 ... \mu_k} \psi_\pm = 0~,
\end{equation} 
for any $k$ even, since ${\bf P}_\pm^t = {\bf P}_\mp$ and ${\bf P}_\pm \Gamma_\mu = \Gamma_\mu {\bf P}_\mp$. For $\psi_\pm$ and $\chi_\pm$ fermionic, two useful Fierz identities are 
\begin{align}\label{eq:10dfierz}
\psi_\pm \, {\overline \chi_\pm} &= - \tfrac{1}{32} \left( 2 ( {\overline \chi_\pm} \Gamma^\mu \psi_\pm ) \Gamma_\mu - \tfrac{1}{3} ( {\overline \chi_\pm} \Gamma^{\mu\nu\rho} \psi_\pm ) \Gamma_{\mu\nu\rho} + \tfrac{1}{5!} ( {\overline \chi_\pm} \Gamma^{\mu\nu\rho\sigma\tau} \psi_\pm ) \Gamma_{\mu\nu\rho\sigma\tau} \right) {\bf P}_\mp \nonumber \\
\psi_\pm \, {\overline \chi_\mp} &= - \tfrac{1}{16} \left( ( {\overline \chi_\mp} \psi_\pm ) {\bf 1} - \tfrac{1}{2} ( {\overline \chi_\mp} \Gamma^{\mu\nu} \psi_\pm ) \Gamma_{\mu\nu} + \tfrac{1}{4!} ( {\overline \chi_\mp} \Gamma^{\mu\nu\rho\sigma} \psi_\pm ) \Gamma_{\mu\nu\rho\sigma} \right) {\bf P}_\pm~.
\end{align}
(For either $\psi_\pm$ or $\chi_\pm$ bosonic, \eqref{eq:10dfierz} just acquires an extra minus sign.) The bilinear ${\overline \chi_\pm} \Gamma_{\mu\nu\rho\sigma\tau} \psi_\pm$ defines a five-form that is self-dual if the spinors have positive chirality and anti-self-dual if the spinors have negative chirality. 

For a non-zero bosonic Majorana-Weyl spinor $\epsilon$ with positive chirality, \eqref{eq:10dbilnears} and \eqref{eq:10devenbilnears} imply only the bilinears $\xi_\mu = {\overline \epsilon} \Gamma_\mu \epsilon$ and $\zeta_{\mu\nu\rho\sigma\tau} = {\overline \epsilon} \Gamma_{\mu\nu\rho\sigma\tau} \epsilon$ built from $\epsilon$ do not vanish identically. The Fierz identity \eqref{eq:10dfierz} for $\epsilon$ reads
\begin{equation}\label{eq:10dfierzepsilon}
\epsilon {\overline \epsilon} = \tfrac{1}{32} ( 2 {\slashed \xi} + {\slashed \zeta} ) {\bf P}_-~,
\end{equation}
and a useful subsidiary identity is
\begin{equation}\label{eq:10dfierzepsilon2}
- \Gamma^{\mu\nu} \epsilon {\overline \epsilon} \, \Gamma_\nu = \left[ \left(  \epsilon {\overline \epsilon} - \half {\slashed\xi} \right) \Gamma^\mu + \xi^\mu {\bf 1} \right] {\bf P}_+~.
\end{equation}
The self-dual $5$-form $\zeta = {*\zeta}$ obeys identically $\xi^\mu \zeta_{\mu\nu\rho\sigma\tau} =0$ implying that $\zeta_{\mu\nu\rho\sigma\tau} = 5\, \xi_{[\mu} \Omega_{\nu\rho\sigma\tau]}$ for some $4$-form $\Omega$ on $\RR^{1,9}$ obeying $\xi^\mu \Omega_{\mu\nu\rho\sigma} =0$.

The isotropy algebra $\fh_\epsilon$ of $\epsilon$ is isomorphic to $\fso(7) \ltimes \RR^8 < \fso(1,9)$ and $\fk_\epsilon \cong ( \fso(8) \, / \, \fso(7)  \ltimes \RR^8 ) \oplus \RR$. Relative to a null orthonormal basis $( {\bf e}_+ , {\bf e}_- , {\bf e}_a )$ on $\RR^{1,9}$, with $\eta_{+-} = 1$ and $\eta_{ab} = \delta_{ab}$ on $\RR^8 \subset \RR^{1,9}$, let us fix $\xi = {\bf e}_+$ so that $\Gamma_+ \epsilon =0$. The orientation tensor on $\RR^8$ is $\varepsilon_{a_1 ... a_8}$ with $\varepsilon_{1...8} = -\varepsilon_{+-1...8} = -1$. The components $\Omega_{abcd} = {\overline \epsilon} \Gamma_{abcd} \Gamma_- \epsilon$ define a self-dual $\fso(7)$-invariant Cayley form on $\RR^8$ and $\Gamma_{ab} \epsilon = -\tfrac{1}{6} \Omega_{abcd} \Gamma^{cd} \epsilon$. The quadratic identity $\Omega_{abef} \Omega^{cdef} + 4 \, \Omega_{ab}{}^{cd} - 12 \, \delta_{[a}^{c} \delta_{b]}^d =0$ for the Cayley form permits the decomposition of $\fso(8)$ into  $\fso(7) \oplus \fso(8) \, / \, \fso(7)$ via the respective projection operators $\tfrac{1}{8} ( 6\, \delta_{[a}^{c} \delta_{b]}^d + \Omega_{ab}{}^{cd} )$ and $\tfrac{1}{8} ( 2\, \delta_{[a}^{c} \delta_{b]}^d - \Omega_{ab}{}^{cd} )$. Adapting this basis to a local frame on $\eM$ allows us to express 
\begin{equation}\label{eq:10dtorsion}
\nabla_\mu \epsilon = a_\mu \epsilon + b_\mu^a \Gamma_a \Gamma_-\epsilon + \half \, c_\mu^{ab} \Gamma_{ab} \epsilon~,
\end{equation}
in terms of real one-forms $a$, $b$ and $c$ valued respectively in the $\RR$, $\RR^8$ and $\fso(8) \, / \, \fso(7)$ factors of $\fk_\epsilon$, corresponding to the intrinsic torsion associated with the $\fso(7) \ltimes \RR^8$-structure. Notice that $\nabla_\mu \xi_\nu$ does not involve $c$. Demanding that $\xi$ be a conformal Killing vector fixes $a_- =0$, $a^a = - b_-^a$, $b_+^a =0$ and $b_{ab} + b_{ba} = \delta_{ab} a_+$. Consequently, from \eqref{eq:10dtorsion}, one derives the twistor spinor equation $\eD_\mu \epsilon = \tfrac{1}{10} \Gamma_\mu {\slashed \eD} \epsilon$ with respect to connection $\eD_\mu = \nabla_\mu + {\sf t}_\mu$, where ${\sf t}_\mu = B_{\mu a} \, \Gamma^a \Gamma_- + \half \, C_\mu^{ab} \Gamma_{ab}$. The component $B_{\mu a}$ has $B_{+a} = 0 = B_{-a}$ and $B_{ba} = - \tfrac{3}{4} b_{[ba]} - \tfrac{1}{8} \Omega_{bacd} b^{cd}$, whence spanning $\fso(7) < \fh_\epsilon$. The other component $C_\mu^{ab}$ has $C_+^{ab} = - c_+^{ab} + \tfrac{1}{8} b^{[ab]} - \tfrac{1}{16} \Omega^{abcd} b_{cd}$, $C_-^{ab} = - c_-^{ab}$, $C_c^{ab} = - c_c^{ab} + \half \delta_c^{[a} a^{b]} - \tfrac{1}{4} \Omega_{c}{}^{abd} a_d$, whence spanning $T^* \eM \otimes \fso(8) \, / \, \fso(7)$. 

From equation (3.34) in \cite{Bergshoeff:1982az}, for bosonic supersymmetric backgrounds, setting to zero the full Poincar\'{e} plus conformal supersymmetry variations of the gravitino and dilatino in the minimal conformal supergravity multiplet in ten dimensions also yields a twistor spinor equation. This follows after performing a Weyl transformation with parameter $\varphi^{10/w}$, in terms of the dilaton $\varphi$ that is defined in \cite{Bergshoeff:1982az} with Weyl weight $w$. The twistor spinor equation from the gravitino variation is with respect to a connection with three-form torsion proportional to $\varphi^{54/w}$ times the Hodge-dual of the seven-form flux for the six-form gauge field in the conformal supergravity multiplet. This can be identified with the connection $\eD$ above only when ${\sf t}$ can be related to components of the three-form torsion. However, even when this can be done, the identification must also be compatible with the additional constraints which come from setting to zero the supersymmetry variation of the dilatino in the conformal supergravity background.     

\subsection{Yang-Mills supermultiplet}
\label{sec:10dSYMonshell} 

The on-shell Yang-Mills supermultiplet in ${\sf d}=10$ contains a bosonic gauge field $A_\mu$ and a fermionic Majorana-Weyl spinor $\lambda$ (we take $\lambda$ with positive chirality, i.e. $\Gamma \lambda = \lambda$). 

The supersymmetry transformations on $\RR^{1,9}$ are     
\begin{align}\label{eq:10dsusy}
\delta_\epsilon A_\mu &= {\overline \epsilon} \Gamma_\mu \lambda \nonumber \\
\delta_\epsilon \lambda &= -\tfrac{1}{2} F^{\mu\nu} \Gamma_{\mu\nu} \epsilon~,
\end{align}
where $\epsilon$ is a constant bosonic Majorana-Weyl spinor with positive chirality. Squaring \eqref{eq:10dsusy} gives $\delta_\epsilon^2 = \xi^\mu \partial_\mu + \delta_\Lambda$ with gauge parameter $\Lambda = - \xi^\mu A_\mu$, after imposing the fermionic equation of motion ${\slashed D} \lambda =0$. Whence, the supersymmetry algebra closes on-shell. Under \eqref{eq:10dsusy}, one finds $\delta_\epsilon ( {\slashed D} \lambda ) = ( D^\nu F_{\mu\nu} ) \Gamma^\mu \epsilon$, where $D^\nu F_{\mu\nu} = 0$  is the equation of motion for $A_\mu$.

Up to boundary terms, the lagrangian
\begin{equation}\label{eq:10dsusylag}
\eL_{\mathrm{SYM}} = -\tfrac{1}{4} ( F_{\mu\nu} , F^{\mu\nu} ) -\half ( {\overline \lambda} , {\slashed D} \lambda )~.
\end{equation}
is invariant under \eqref{eq:10dsusy}, and gives $D^\nu F_{\mu\nu} =0$ and ${\slashed D} \lambda = 0$ as equations of motion.

We now consider the minimally coupled version of \eqref{eq:10dsusy} on a ten-dimensional lorentzian spin manifold $\eM$. Notice that they are automatically Weyl-covariant on $\eM$. Squaring them gives  
\begin{align}\label{eq:10dsusysquared}
\delta_\epsilon^2 A_\mu =& \cL_\xi A_\mu + D_\mu \Lambda \nonumber \\
\delta_\epsilon^2 \lambda =& \xi^\mu \nabla_\mu \lambda + [ \lambda , \Lambda ] - \Gamma^{\mu\nu} \epsilon ( \nabla_\mu {\overline \epsilon} ) \Gamma_\nu \lambda + \left( \epsilon {\overline \epsilon} - \half {\slashed \xi} \right) {\slashed D} \lambda \nonumber \\
=& \cL_\xi \lambda + [ \lambda , \Lambda ] + \left( \epsilon {\overline \epsilon} - \half {\slashed \xi} \right) {\slashed D} \lambda \nonumber\\
&+ \tfrac{1}{32} \left[ 9\, (\nabla_\mu \xi^\mu ) \, {\bf 1} - \left( \nabla_\mu \xi_\nu - 5\, {\overline \epsilon} \, \Gamma_{\mu\nu\rho} \nabla^\rho \epsilon \right) \Gamma^{\mu\nu} - \left( {\overline \epsilon} \, \Gamma_{\mu\nu\rho} \nabla_\sigma \epsilon - \tfrac{1}{24} \, \nabla^\tau \zeta_{\mu\nu\rho\sigma\tau} \right) \, \Gamma^{\mu\nu\rho\sigma} \right] \lambda~.
\end{align}
Determining the general conditions necessary for on-shell closure on $\eM$ is complicated by the possibility of additional terms modifying the fermion equation of motion. We will therefore not attempt to address this problem systematically. However, we shall obtain an interesting class of solutions by focussing on the following ansatz. 

Consider an $\epsilon$ which obeys the two conditions 
\begin{equation}\label{eq:10dTspinor}
\nabla_\mu \epsilon = \tfrac{1}{8} H_{\mu\nu\rho} \Gamma^{\nu\rho} \epsilon \; , \quad\quad G_\mu \Gamma^\mu \epsilon = \tfrac{1}{12} H_{\mu\nu\rho} \Gamma^{\mu\nu\rho} \epsilon~,
\end{equation}
for some three-form $H$ and one-form $G$ on $\eM$. The first condition implies $\nabla_\mu \xi_\nu = \half \, H_{\mu\nu\rho} \xi^\rho$, whence $\xi$ is a Killing vector. Hitting the second condition in \eqref{eq:10dTspinor} with ${\overline \epsilon}$ implies $\xi^\mu G_\mu = 0$. Hitting it with ${\overline \epsilon} \Gamma_{\alpha\beta}$ and ${\overline \epsilon} \Gamma_{\alpha\beta\gamma\delta}$ gives two more useful subsidiary identifies
\begin{align}\label{eq:10dTspinoridentity}
\tfrac{1}{6} H^{\mu\nu\rho} \zeta_{\alpha\beta\mu\nu\rho} &= H_{\alpha\beta\mu} \xi^\mu - 4\, G_{[\alpha} \xi_{\beta]} \nonumber \\
H_{\mu\nu[\alpha} \zeta_{\beta\gamma\delta]}^{\quad\;\;\;\; \mu\nu} &= 2\, H_{[\alpha\beta\gamma} \xi_{\delta]} - \zeta_{\alpha\beta\gamma\delta\mu} G^\mu~.
\end{align}
In fact,  $\xi^\mu G_\mu = 0$ and \eqref{eq:10dTspinoridentity} are equivalent to the second condition in \eqref{eq:10dTspinor}. 

Substituting \eqref{eq:10dTspinor} into \eqref{eq:10dsusysquared} and using \eqref{eq:10dTspinoridentity} gives
\begin{align}\label{eq:10dsusysquared2}
\delta_\epsilon^2 A_\mu &= \cL_\xi A_\mu + D_\mu \Lambda \nonumber \\
\delta_\epsilon^2 \lambda &= \cL_\xi \lambda + [ \lambda , \Lambda ] + \left[ \epsilon {\overline \epsilon} - \half {\slashed \xi} \right] \fF~,
\end{align} 
where
\begin{equation}\label{eq:10dTspinorpsieom}
\fF = \left( {\slashed D} - {\slashed G} - \tfrac{1}{4} {\slashed H} \right) \lambda~.
\end{equation}
Thus on-shell closure of the supersymmetry algebra is established for any $\epsilon$ obeying \eqref{eq:10dTspinor}, with $\fF = 0$ in \eqref{eq:10dTspinorpsieom} describing the equation of motion for $\lambda$. The transformation of \eqref{eq:10dTspinorpsieom} under \eqref{eq:10dsusy} gives
\begin{equation}\label{eq:10dTspinorpsieomsusyvar}
\delta_\epsilon \fF = \fB_\mu \Gamma^\mu \epsilon~,
\end{equation} 
where
\begin{equation}\label{eq:10dTspinorAeom}
\fB_\mu = D^\nu F_{\mu\nu} - 2 F_{\mu\nu} G^\nu - \half H_{\mu\nu\rho} F^{\nu\rho}~.
\end{equation}
The equation of motion for $A_\mu$ is therefore $\fB_\mu = 0$ and the on-shell identity $D^\mu \fB_\mu = 0$ constrains the data in \eqref{eq:10dTspinor} such that
\begin{equation}\label{eq:10dTspinordV}
2 \nabla_{[\mu} G_{\nu]} = - \half \nabla^\rho H_{\mu\nu\rho} + H_{\mu\nu\rho} G^\rho~.
\end{equation}

If $G = d \Phi$, \eqref{eq:10dTspinordV} implies $d \left( {\mbox{e}}^{-2\Phi} \, {*H} \right) =0$ while \eqref{eq:10dTspinor} and \eqref{eq:10dTspinoridentity} imply 
\begin{align}\label{eq:10dTspinoridentityder}
\nabla_{[\alpha} \left( {\mbox{e}}^{-2\Phi} \xi_{\beta]} \right) &= \tfrac{1}{12} \, {\mbox{e}}^{-2\Phi} H^{\mu\nu\rho} \zeta_{\alpha\beta\mu\nu\rho} \\ \nonumber
\nabla^{\mu} \left( {\mbox{e}}^{-2\Phi} \zeta_{\mu\alpha\beta\gamma\delta} \right) &= -4\, {\mbox{e}}^{-2\Phi} H_{[\alpha\beta\gamma} \xi_{\delta]}~.
\end{align}
In that case, the equations of motion \eqref{eq:10dTspinorpsieom} and \eqref{eq:10dTspinorAeom} follow from the lagrangian
\begin{equation}\label{eq:10dsusylagcurved}
{\mbox{e}}^{-2\Phi} \left[ -\tfrac{1}{4} ( F_{\mu\nu} , F^{\mu\nu} ) -\half ( {\overline \lambda} , {\slashed D} \lambda ) + \tfrac{1}{8} ( {\overline \lambda} , {\slashed H} \lambda ) + \half H^{\mu\nu\rho} ( A_\mu , \partial_\nu A_\rho + \tfrac{1}{3} [ A_\nu , A_\rho ] ) \right]~,
\end{equation}
Moreover, up to boundary terms, \eqref{eq:10dsusylagcurved} is invariant under the minimal coupling of the supersymmetry transformations in \eqref{eq:10dsusy}. The prefactor ${\mbox{e}}^{-2\Phi}$ acts as an effective gauge coupling in \eqref{eq:10dsusylagcurved}. For generic backgrounds with $H \neq 0$, notice that supersymmetry necessitates both a mass term for $\lambda$ and a Chern-Simons coupling for the gauge field. Closure of ${\mbox{e}}^{-2\Phi} {*H}$ ensures that the Chern-Simons coupling is gauge-invariant.  

Before concluding with a brief discussion of supergravity backgrounds and decoupling limits for the on-shell theory above, it is perhaps worth making a few remarks about the novel (partially) off-shell formulation of supersymmetric Yang-Mills theory on $\RR^{1,9}$ that was found by Berkovits in \cite{Berkovits:1993zz} (see also \cite{Evans:1994cb,Baulieu:2007ew}). To match the $16$ off-shell fermionic degrees of freedom of $\lambda$, the $9$ off-shell degrees of freedom of $A_\mu$ are supplemented by $7$ bosonic auxiliary scalar fields $\sigma_i$, where $i=1,...,7$. The supersymmetry parameter $\epsilon$ is also supplemented by seven linearly independent bosonic Majorana-Weyl spinors $\nu_i$, each with the same positive chirality as $\epsilon$. The index $i$ corresponds to the vector representation of the $\fso(7)$ factor in the isotropy algebra of $\epsilon$.

The supersymmetry transformations on $\RR^{1,9}$ are 
\begin{align}\label{eq:10dsusyoffshell}
\delta_{\epsilon , \nu} A_\mu &= {\overline \epsilon} \Gamma_\mu \lambda \nonumber \\
\delta_{\epsilon , \nu} \lambda &= -\tfrac{1}{2} F^{\mu\nu} \Gamma_{\mu\nu} \epsilon + \sigma_i \nu_i \nonumber \\
\delta_{\epsilon , \nu} \sigma_i &= {\overline \nu_i}  {\slashed D} \lambda~.
\end{align}
Squaring \eqref{eq:10dsusyoffshell} gives $\delta_{\epsilon , \nu}^2 = \xi^\mu \partial_\mu + \delta_\Lambda$ off-shell, via the identities
\begin{equation}\label{eq:10doffshellnuidentity}
{\overline \epsilon} \Gamma_\mu \nu_i = 0 \; , \quad\quad {\overline \nu_i} \Gamma_\mu \nu_j = \delta_{ij} \, \xi_\mu \; , \quad\quad \epsilon {\overline \epsilon} + \nu_i {\overline \nu_i} = \half {\slashed \xi}~.
\end{equation}
Furthermore, up to boundary terms, the lagrangian
\begin{equation}\label{eq:10dsusylagoffshell}
-\tfrac{1}{4} ( F_{\mu\nu} , F^{\mu\nu} ) -\half ( {\overline \lambda} , {\slashed D} \lambda ) + \half ( \sigma_i , \sigma_i )~,
\end{equation}
is invariant under \eqref{eq:10dsusyoffshell}. The action of $\fk_\epsilon$ on $\epsilon$ and $\nu_i$ generates a nine-dimensional subspace of the sixteen-dimensional vector space of Majorana-Weyl spinors on $\RR^{1,9}$. Whence, this formulation is manifestly off-shell with respect to nine of the sixteen linearly independent supercharges in Minkowski space.  

On a curved ten-dimensional lorentzian spin manifold $\eM$, the minimal coupling of \eqref{eq:10dsusyoffshell} is not automatically Weyl-covariant. However, the supersymmetry transformations can be conformally coupled by adding an improvement term $- \tfrac{3}{5}\, {\overline \lambda} {\slashed \nabla} \nu_i$ to the right hand side of $\delta_{ \epsilon , \nu} \sigma_i$, relative to the minimal coupling of \eqref{eq:10dsusyoffshell} on $\eM$. If $\epsilon$ and $\nu_i$ are twistor spinors, one finds that $\delta_{\epsilon , \nu}^2 = \cL_\xi + \delta_\sigma + \delta_\Lambda + \delta_\rho$ off-shell, where $\delta_\sigma$ is a Weyl variation with parameter $\sigma = -\tfrac{1}{10} \nabla_\mu \xi^\mu$ and $\delta_\rho$ is an infinitesimal $\fso(7)$ rotation with parameter $\rho_{ij} = \tfrac{4}{5} ( {\overline \nu_i} {\slashed \nabla} \nu_j - {\overline \nu_j} {\slashed \nabla} \nu_i )$, provided the fields $( A_\mu , \lambda , \sigma_i )$ are assigned their canonical Weyl weights $(0,-\tfrac{3}{2},-2)$. 

The integral on $\eM$ of the minimal coupling of lagrangian \eqref{eq:10dsusylagoffshell} is neither Weyl-invariant nor supersymmetric for generic twistor spinors $\epsilon$ and $\nu_i$. Even for the subclass of algebraic twistor spinors, which yielded rigid supersymmetric lagrangian gauge theories in lower dimensions, one finds that there are no generic solutions which can preserve the off-shell rigid supersymmetry in ten dimensions. 

\subsubsection{Supergravity backgrounds}
\label{sec:sugrabackgrounds} 

The data $( \eM , g,G=d\Phi ,H=dB)$ solving \eqref{eq:10dTspinor} correspond to bosonic supersymmetric backgrounds of minimal Poincar\'{e} supergravity in ten dimensions (with dilaton $\Phi$ and $2$-form gauge potential $B$). From this perspective, the first and second equations in \eqref{eq:10dTspinor} correspond respectively to the vanishing of the supersymmetry variations of the gravitino and dilatino. A classification of all such backgrounds which solve the supergravity equations of motion was obtained in \cite{Gran:2005wf,Gran:2007kh,Papadopoulos:2009br} (as a subclass of the supersymmetric solutions of heterotic supergravity wherein the gauge sector is set to zero). 

Majorana-Weyl spinors in ${\sf d}=10$ can be thought of locally as vectors in $\RR^{16}$ so there can be no more than sixteen linearly independent real $\epsilon$ which solve \eqref{eq:10dTspinor}.  For a given background, let $\eN \leq 16$ denote the maximum number of linearly independent positive-chirality Majorana-Weyl spinors solving \eqref{eq:10dTspinor}.

Let us now briefly survey of the landscape of $\eN \geq 8$ solutions. As was first proven in \cite{FigueroaO'Farrill:2002ft}, the only $\eN =16$ solution is $\eM = \RR^{1,9}$ with $G=0$ and $H=0$. Between $8 < \eN < 16$, there are only the $\eN = 10,12,14$ \lq parallelisable' plane wave solutions found in \cite{FigueroaO'Farrill:2003qm} with $G=0$. The $\eN =8$ solutions are classified in \cite{Papadopoulos:2008rx} and fall into three classes. In the first class, $\eM$ is a lorentzian group manifold corresponding to one of the $\eN =8$ entries in table 4 of \cite{FigueroaO'Farrill:2003qm}. In the second class, $\eM$ is a principle bundle ${\mbox{P}}(\fG ,X)$, equipped with a connection $C$. The base $X$ is a hyperK\"{a}hler four-manifold, on which the curvature of $C$ must be anti-self-dual. The fibre group $\fG$ must have a six-dimensional lorentzian self-dual lie algebra which, via the results of \cite{Chamseddine:2003yy}, restricts $\fG$ to be locally isometric to either $\RR^{1,5}$, $\AdS_3 \times S^3$ or a plane wave. If $C$ is flat, $\eM$ is locally isometric to either $\fG \times X$ or a generalised five-brane solution with $\fG$ as its worldvolume and $X$ as its transverse space. In the third class, $\eM$ takes the form of a superposition of fundamental string and pp-wave solutions together with a null rotation.  

\subsubsection{Decoupling limit of Chapline-Manton theory}
\label{sec:DecouplingCM} 

The theory constructed by Chapline and Manton in \cite{Chapline:1982ww} describes the minimally supersymmetric coupling of Yang-Mills and Poincar\'{e} gravity supermultiplets in ten dimensions. This provides a local on-shell representation of the supersymmetry algebra. We will now conclude with a brief explanation how rigid supersymmetric Yang-Mills theory on a supergravity background of the kind described above can be recovered from a particular decoupling limit of the Chapline-Manton theory. 

In the notation of \cite{Chapline:1982ww} (with appended "CM" subscripts) we first set the gravitino $\psi_{\mathrm{CM}}$ and dilatino $\lambda_{\mathrm{CM}}$ equal to zero. Now define $\Phi := \tfrac{3}{8} \, {\mbox{ln}} \, \phi_{\mathrm{CM}}$, $\lambda := {\mbox{e}}^{\Phi} {\sf g}_{{\sf YM}} \chi_{\mathrm{CM}}$, $H := 3\sqrt{2} \, {\mbox{e}}^{-2\Phi} \kappa f_{\mathrm{CM}}$ in terms of their dilaton $\phi_{\mathrm{CM}}$, gaugino $\chi_{\mathrm{CM}}$ and three-form field $f_{\mathrm{CM}}$, where $\kappa$ and ${\sf g}_{{\sf YM}}$ correspond to the gravitational and Yang-Mills coupling constants that we find it convenient to append to their expressions. To recover a supersymmetric background for the supergravity fields, the supersymmetry variations of $\psi_{\mathrm{CM}}$ and $\lambda_{\mathrm{CM}}$ must also be set to zero. However, this can only be done whilst keeping the fields the Yang-Mills supermultiplet dynamical in the $\kappa / {\sf g}_{{\sf YM}} \rightarrow 0$ limit. This is tantamount to taking the Planck mass to infinity whereupon the gravitational dynamics become decoupled. In this limit, one recovers precisely \eqref{eq:10dTspinor} and \eqref{eq:10dsusy} from the supersymmetry transformations in equation (11) of \cite{Chapline:1982ww}. The lagrangian \eqref{eq:10dsusylagcurved} corresponds to the $\kappa$-independent contribution to (10) in \cite{Chapline:1982ww} that is subleading in the $\kappa / {\sf g}_{{\sf YM}} \rightarrow 0$ limit.


\section*{Acknowledgments}

I would like to thank Mohab Abou Zeid, Jos\'{e} Figueroa-O'Farrill and Stefan Hollands for some helpful discussions at various stages of this work. The financial support provided by ERC Starting Grant QC \& C 259562 is gratefully acknowledged.

\bibliographystyle{utphys}
\bibliography{CurvedSUSY}

\end{document}